\documentclass[aps,prd,onecolumn,groupedaddress,showpacs,nofootinbib,amssymb]{revtex4}
\usepackage[dvips]{graphicx}
\usepackage{amssymb}
\usepackage{amsmath}
\usepackage{graphicx,,color}
\usepackage{amsfonts}
\usepackage{bm}
\usepackage{cancel}
\usepackage{comment}

\allowdisplaybreaks[4]

\begin{document}

\tolerance=5000

\title{Rescaled Einstein-Hilbert Gravity: Inflation and the Swampland Criteria}
\author{V.K.~Oikonomou,$^{1,2}$\,\thanks{v.k.oikonomou1979@gmail.com}}
\author{Ifigeneia Giannakoudi,$^{1}$\,\thanks{ifigeneiagiannakoudi@gmail.com}}
\author{Achilles Gitsis,$^{1}$\,\thanks{agitsis62@gmail.com}}
\author{Konstantinos-Rafail Revis,$^{1}$\,\thanks{reviskostis@gmail.com}}
\affiliation{$^{1)}$ Department of Physics, Aristotle University
of Thessaloniki, Thessaloniki 54124,
Greece\\
$^{2)}$ Laboratory for Theoretical Cosmology, Tomsk State
University of Control Systems and Radioelectronics, 634050 Tomsk,
Russia (TUSUR)}

\tolerance=5000

\begin{abstract}
In this work we shall study a class of $f(R,\phi)$ gravity models
which during the inflationary era, which is the large curvature
regime, result to an effective inflationary Lagrangian that
contains a rescaled Einstein-Hilbert term $\alpha R$ in the
presence of a canonical minimally coupled scalar field. The
dimensionless parameter $\alpha$ is chosen to take values in the
range $0<\alpha<1$ and the main motivation for studying these
rescaled Einstein-Hilbert $f(R,\phi)$ gravities, is the fact that
the rescaled action may render an otherwise incompatible canonical
scalar field theory with the Swampland criteria, to be compatible
with the Swampland criteria. As we will show, by studying a large
number of inflationary potentials appearing in the 2018 Planck
collaboration article for the constraints on inflation, the
simultaneous compatibility with both the Planck constraints and
the Swampland criteria, is achieved for some models, and the main
characteristic of the models for which this is possible, is the
small values that the parameter $\alpha$ must take.
\end{abstract}

\maketitle

\section{Introduction}

The next two decades are expected to be hopefully quite fruitful
for astrophysics and cosmology. Several groundbreaking experiments
and missions are expected to commence and yield valuable
information for our Universe's primordial era, like the LISA space
mission \cite{Baker:2019nia,Smith:2019wny} or the DECIGO
\cite{Seto:2001qf,Kawamura:2020pcg}. Specifically, the observation
of primordial gravitational waves directly in experiments, can
verify that the inflationary scenario actually occurred
primordially. Most of the space interferometers and other
experiments will probe scales much smaller than 10$\,$Mpc, where
the Cosmic Microwave Background scalar curvature perturbations
become highly non-linear. Apart from the Square Kilometer Array
(SKA) \cite{Bull:2018lat} and the NANOGrav collaboration
\cite{Arzoumanian:2020vkk,Pol:2020igl}, which may probe primordial
gravitational waves during the last stages of the radiation
domination era, and the Planck collaboration \cite{Akrami:2018odb}
which probes large scales of very small frequencies, the rest of
the experiments and space missions will probe primordial
gravitational waves which are deeply in the radiation domination
era. Thus, the direct observations of the primordial tensor
perturbations will signify that inflation took place primordially.
The B-mode polarization modes can indicate the existence of
primordial tensor perturbations primordially at large scales (low
multipoles of the CMB $\ell \leq 10$), or a direct gravitational
lensing conversion of $E$-modes to $B$-modes at late-time (for
small angular scales in the CMB) \cite{Denissenya:2018mqs}. We
also need to note that primordial B-mode polarization can be
produced in alternative to inflation theories, except from the
initial version of the Ekpyrotic scenario.

The standard description of the inflationary scenario is by using
a canonical scalar field, the so-called inflaton
\cite{Linde:2007fr,Gorbunov:2011zzc,Lyth:1998xn,Martin:2018ycu}.
Although the only fundamental scalar that has ever been observed
is the Higgs particle, scalar fields are inherent features of the
most successful to date fundamental theory of elementary particle
interactions at high energies, namely string theory in its various
forms. Moreover, since inflationary era occurs primordially and
chronically quite close the Planck era, it is possible that the
Planck era might have left its imprint in the inflationary effective
Lagrangian, in terms of some higher order curvature terms, that
may deviate from the linear curvature term of Einstein-Hilbert
gravity. Modified gravity theory in its various forms, serves as
the most appealing candidate class of theories which contains
higher order curvature corrections
\cite{reviews1,reviews2,reviews3,reviews4,reviews5,reviews6}. The
most appealing frameworks unify inflation and the late-time
acceleration eras, see for example the pioneer work
\cite{Nojiri:2003ft} and Refs.
\cite{Nojiri:2007as,Nojiri:2007cq,Cognola:2007zu,Nojiri:2006gh,Appleby:2007vb,Elizalde:2010ts,Odintsov:2020nwm,Oikonomou:2020qah,Oikonomou:2020oex}
for later developments along the unification of cosmological eras
line of research. In view of the modified gravity perspective for
describing the effective inflationary Lagrangian, in this work we
shall extensively study the effects of a class of $f(R,\phi)$
gravity theories which during the inflationary era are described
by a rescaled canonical scalar field theory, with the
Einstein-Hilbert curvature term being replaced by $\sim \alpha R$.
These models were developed and firstly appeared in
\cite{Oikonomou:2020qah} and further studied in
\cite{Oikonomou:2020oex}. The prominent feature of the rescaled
Einstein-Hilbert gravity is that it is possible to reconcile the
Swampland conjectures
\cite{Vafa:2005ui,Ooguri:2006in,Palti:2020qlc,Mizuno:2019bxy,Brandenberger:2020oav,Blumenhagen:2019vgj,Wang:2019eym,Benetti:2019smr,Palti:2019pca,Cai:2018ebs,Akrami:2018ylq,Mizuno:2019pcm,Aragam:2019khr,Brahma:2019mdd,Mukhopadhyay:2019cai,Brahma:2019kch,Haque:2019prw,Heckman:2019dsj,Acharya:2018deu,Elizalde:2018dvw,Cheong:2018udx,Heckman:2018mxl,Kinney:2018nny,Garg:2018reu,Lin:2018rnx,Park:2018fuj,Olguin-Tejo:2018pfq,Fukuda:2018haz,Wang:2018kly,Ooguri:2018wrx,Matsui:2018xwa,Obied:2018sgi,Agrawal:2018own,Murayama:2018lie,Marsh:2018kub,Storm:2020gtv,Trivedi:2020wxf,Sharma:2020wba,Odintsov:2020zkl,Mohammadi:2020twg,Trivedi:2020xlh,Han:2018yrk,Achucarro:2018vey,Akrami:2020zfz,Oikonomou:2020oex},
see also \cite{Colgain:2018wgk,Colgain:2019joh,Banerjee:2020xcn},
with canonical scalar field theory. In the present we shall extend
the study \cite{Oikonomou:2020oex} by examining a large number of
inflationary potentials, and confronting the resulting models with
the Planck data for the rescaled versions of the canonical scalar
field theories. We also discuss when the Swampland conjectures are
satisfied for the resulting rescaled models. This work is
organized as follows: In section II we present the theoretical
framework needed in the case of $\alpha R$, both for the
inflationary and the Swampland part, in section III we test the
compliance of several inflationary models with the latest Planck
data.  In section IV we complete our analysis by testing the
validity of the Swampland criteria for some of the models of
section III. Finally the conclusions follow at the end of the
article.

Before starting, let us note that for the background metric we
shall choose a flat Friedmann Robertson Walker (FRW) metric, with
the line element given by,
\begin{equation}
    \centering\label{frw}
    d s^2 = - d t^2 + a(t) \sum_{i = 1}^3 d x_i^2,
\end{equation}
where $a(t)$ is the scale factor as usual. For the flat FRW
metric, the non-zero components of the Ricci tensor are,
\begin{equation}
    \centering\label{r00}
    R_{0 0 } = - 3 \frac{\ddot{a}}{a},
\end{equation}
\begin{equation}
    \centering\label{rii}
    R_{1 1} = R_{2 2} = R_{3 3} = \ddot{a}a + 2 \dot{a}^2,
\end{equation}
while the Ricci scalar is given by,
\begin{equation}
    R = 6 \dot{H} + 12 H^2,
\end{equation}
where $H = \frac{\dot{a}}{a}$ is the Hubble parameter. Moreover,
hereafter $\kappa=\frac{1}{M_p}$, where $M_p$ is the reduced
Planck mass, and we shall use natural units $c = \hbar = 1$
physical dimensions system.

\section{The $f(R)$ Gravity Scalar Field Model, Inflationary Effective Theory and the Swampland Criteria}

The rescaled Einstein-Hilbert gravity models originate from the
following action \cite{Oikonomou:2020oex},

The $f(R)$ gravity scalar field model action has the following
form,
\begin{equation}
\label{action} \centering
\mathcal{S}=\int{d^4x\sqrt{-g}\left(\frac{f(R)}{2\kappa^2}-\frac{1}{2}\partial_\mu\phi\partial^\mu\phi-V(\phi)\right)}\,
,
\end{equation}
with the $f(R)$ gravity being of the form,
\begin{equation}\label{frini}
f(R)=R-\gamma  \lambda  \Lambda -\lambda  R \exp
\left(-\frac{\gamma  \Lambda }{R}\right)-\frac{\Lambda
\left(\frac{R}{m_s^2}\right)^{\delta }}{\zeta }\, .
\end{equation}
As it was shown in Ref. \cite{Oikonomou:2020oex}, the above $f(R)$
gravity at leading order in the large curvatures limit takes the
form,
\begin{equation}\label{expapprox}
\lambda  R \exp \left(-\frac{\gamma  \Lambda }{R}\right)\simeq
-\gamma \lambda  \Lambda -\frac{\gamma ^3 \lambda \Lambda^3}{6
R^2}+\frac{\gamma ^2 \lambda  \Lambda ^2}{2 R}+\lambda  R\, ,
\end{equation}
therefore, the effective inflationary action during the
inflationary era is at leading order in the large curvature limit
$R\to \infty$,
\begin{equation}\label{effectiveaction}
\mathcal{S}=\int
d^4x\sqrt{-g}\left(\frac{1}{2\kappa^2}\left(\alpha R+ \frac{\gamma
^3 \lambda \Lambda ^3}{6 R^2}-\frac{\gamma ^2 \lambda \Lambda
^2}{2 R}-\frac{\Lambda}{\zeta
}\left(\frac{R}{m_s^2}\right)^{\delta
}+\mathcal{O}(1/R^3)+...\right)-\frac{1}{2}\partial_\mu\phi\partial^\mu\phi-V(\phi)\right)\,
,
\end{equation}
where $\alpha=1-\lambda$. The gravitational action
(\ref{effectiveaction}) is the rescaled Einstein-Hilbert canonical
scalar field theory, and it will be the starting point of our
analysis. Let us note that the action (\ref{effectiveaction}) is a
Jordan frame effective field theory, and also let us note that the
Swampland criteria apply not the rescaled effective action
(\ref{effectiveaction}), but to the original action
(\ref{action}), see Ref. \cite{Oikonomou:2020oex} for a detailed
explanation of this. The action (\ref{effectiveaction}) is at
leading order written as follows,
\begin{equation}
    \centering\label{act}
    S = \int d^4 x \sqrt{- g}\left(\frac{\alpha R}{2 \kappa^2} - \frac{1}{2}g^{\mu \nu} \partial_\mu \phi \partial_\nu \phi - V(\phi) \right),
\end{equation}
where $\alpha$ is a dimensionless parameter which will be assumed
to take values in the range $0 \leq \alpha \leq 1$. The scalar
field, satisfies the following equation of motion,
\begin{equation}
    \centering\label{eomphi}
    \ddot{\phi} + 3 H \dot{\phi} + V' = 0,
\end{equation}
where the ``prime'' denotes differentiation with respect to
$\phi$. Thus by varying the original action (\ref{action}) with
respect to the metric, for the model at hand, we obtain at leading
order,
\begin{equation}
    \centering\label{eom}
    \frac{\alpha}{\kappa^2}\left(R_{\mu \nu} - \frac{1}{2}R g_{\mu \nu}\right) = \partial_\mu \phi \partial_\nu \phi - g_{\mu \nu}\left(\frac{1}{2}g^{\rho \sigma}\partial_\rho \phi \partial_\sigma \phi +V(\phi) \right).
\end{equation}
For the FRW metric, the Einstein field equations (\ref{eom}) yield
the Friedmann equations,
\begin{equation}
    \centering\label{fr1}
    \frac{3\alpha}{\kappa^2} H^2 =\frac{1}{2}\dot{\phi}^2 + V(\phi),
\end{equation}
\begin{equation}
    \centering\label{fr2}
    \frac{2 \alpha}{\kappa^2}\dot{H} =-\dot{\phi}^2.
\end{equation}
The slow-roll condition for the inflationary era is,
\begin{equation}
    \centering\label{inf}
    V(\phi) \gg \dot{\phi}^2,
\end{equation}
and in conjunction with (\ref{fr1}) and (\ref{fr2}),  it takes the
form,
\begin{equation}
    \centering\label{cond}
    \frac{|\dot{H}|}{H^2} \ll 1.
\end{equation}
Due to (\ref{cond}), the first slow-roll parameter is defined as
\begin{equation}
    \centering\label{eps1}
     \epsilon_1 = -\frac{\dot{H}}{H^2},
\end{equation}
and (\ref{fr1}) becomes
\begin{equation}
    \centering\label{h2}
    H^2 = \frac{\kappa^2}{3 \alpha}V(\phi).
\end{equation}
The condition for the duration of the inflationary era to be
sufficiently long is given by,
\begin{equation}
    \centering\label{time}
    |\ddot{\phi}| \ll 3 H |\dot{\phi}|,
\end{equation}
and the second slow-roll parameter is defined as,
\begin{equation}
    \centering\label{eps2}
     \epsilon_2 = \frac{\ddot{\phi}}{H \dot{\phi}}.
\end{equation}
Due to (\ref{time}), from (\ref{eomphi}) we obtain
\begin{equation}
    \centering\label{dphi}
    \dot{\phi} = - \frac{V'}{3 H}\, ,
\end{equation}
and
\begin{equation}
    \centering\label{ddphi}
    \ddot{\phi} = - \frac{\dot{H}}{H}\dot{\phi} -V'' \frac{\dot{\phi}}{3 H}.
\end{equation}
By direct substitution of (\ref{fr2}), (\ref{h2}), (\ref{dphi})
and (\ref{ddphi}) in Eqs. (\ref{eps1}) and (\ref{eps2}), the
slow-roll parameters become,
\begin{equation}
    \centering\label{eps1 to eps}
     \epsilon_1 = \alpha  \epsilon\, ,
\end{equation}
and
\begin{equation}
    \centering\label{eps2 to eta}
     \epsilon_2 = -\alpha \eta +  \epsilon_1,
\end{equation}
where,
\begin{equation}
    \centering\label{eps}
     \epsilon = \frac{1}{2 \kappa^2}\frac{V'^2}{V^2},
\end{equation}
\begin{equation}
    \centering\label{eta}
    \eta = \frac{1}{\kappa^2}\frac{V''}{V} ,
\end{equation}
are the first and second potential slow-roll parameters,
respectively. The $e$-foldings number is defined by,
\begin{equation}
    \centering\label{efold}
    N(\phi) = \int_t^{t_{end}} H d t,
\end{equation}
where $t_{end}$ denotes the end of the inflation, and by using
(\ref{h2}) and (\ref{dphi}), Eq. (\ref{efold}) takes the form,
\begin{equation}
    \centering\label{N}
    N(\phi) = \frac{\kappa^2}{\alpha} \int_{\phi_{end}}^\phi \frac{V}{V'}d \phi,
\end{equation}
where $\phi_{end}$ is the value of the inflaton at the end of the
inflation. The spectral index is defined as,
\begin{equation}
    \centering\label{spec}
    n_s - 1 = -4\epsilon_1 -2\epsilon_2 , 
\end{equation}
and substituting (\ref{eps1 to eps}) and (\ref{eps2 to eta}), it becomes
\begin{equation}
    \centering\label{ns}
    n_s = 1 + 2 \alpha \eta - 6 \alpha  \epsilon.
\end{equation}
Moreover, the tensor-to-scalar ratio, defined as the ratio of the
tensor perturbations to the scalar perturbations, is given by,
\begin{equation}
    \centering\label{ttsr}
    r = 8 \kappa^2 \frac{\dot{\phi}^2}{H^2},
\end{equation}
which in our case reduces to,
\begin{equation}
    \centering\label{r}
    r = 16 \alpha  \epsilon.
\end{equation}
At this point let us recall the Swampland criteria, that firstly
appeared in Refs. \cite{Vafa:2005ui,Ooguri:2006in} and were
further studied in Refs.
\cite{Palti:2020qlc,Mizuno:2019bxy,Brandenberger:2020oav,Blumenhagen:2019vgj,Wang:2019eym,Benetti:2019smr,Palti:2019pca,Cai:2018ebs,Mizuno:2019pcm,Aragam:2019khr,Brahma:2019mdd,Mukhopadhyay:2019cai,Marsh:2019lhu,Brahma:2019kch,Haque:2019prw,Heckman:2019dsj,Acharya:2018deu,Elizalde:2018dvw,Cheong:2018udx,Heckman:2018mxl,Kinney:2018nny,Garg:2018reu,Lin:2018rnx,Park:2018fuj,Olguin-Tejo:2018pfq,Fukuda:2018haz,Wang:2018kly,Ooguri:2018wrx,Matsui:2018xwa,Obied:2018sgi,Agrawal:2018own,Murayama:2018lie,Marsh:2018kub,Storm:2020gtv,Trivedi:2020wxf,Sharma:2020wba,Odintsov:2020zkl,Mohammadi:2020twg,Trivedi:2020xlh},
which are the following using reduced Planck units ($\kappa=1$),
\begin{itemize}
        \item The Swampland Distance Conjecture. It limits the validity of an
effective field theory and sets an upper limit for the maximum
traversable range for a scalar field as it follows:
    \begin{equation}
        \centering\label{S.C. deltaphi}
        \Delta\phi\leq f\sim\mathcal{O}(1).
    \end{equation}
    \item The de Sitter conjecture. It states that it is impossible to
create De Sitter vacua in string theory so, a lower limit for the
gradient of scalar potentials should be applied:
    \begin{equation}
    \centering\label{S.C. v'/v}
    \frac{V'}{V}\geq g\sim\mathcal{O}(1).
    \end{equation}
\end{itemize}
An other useful expression of the Swampland criteria is the following:
  \begin{equation}
    \centering\label{S.C. v''/v}
    \frac{V''}{V}\leq -h\sim\mathcal{O}(1).
    \end{equation}
It is important to underline that $f$, $g$ and $h$ are just some
arbitrary dimensionless constants that do not symbolize a physical
parameter and that we have expressed the Swampland criteria in
reduced Planck units by setting $\kappa^2=1$. If we take into
consideration the definition of $ \epsilon$ provided by
(\ref{eps}) and the de-Sitter conjecture it is obvious that the
first slow-roll parameter should satisfy,
\begin{equation}
    \centering\label{S.C. with epsilon}
    \Big{|}\frac{V'}{V} \Big{|}=\sqrt{2 \epsilon}\geq 1.
\end{equation}
We are able to overcome this obstacle if we recall the fact that
in the slow-roll conditions, the slow-roll indices are quantified
by the conditions  $ \epsilon_1\ll1$ and $ \epsilon_2\ll1$ and not
by the conditions $ \epsilon\ll 1$ and $\eta\ll 1$. The antithesis
between inflationary constraints and the limitations obtained by
the Swampland criteria is obvious at this point. As it is
underlined in Ref. \cite{Oikonomou:2020oex}, we are able to
overcome this obstacle if we recall that $ \epsilon_1$ and
$\epsilon_2$ have to be much smaller than unity, rather than the
potential slow-roll parameters. Bearing this result into mind we
are able to satisfy both the inflationary constraints and the
limitations obtained by the Swampland criteria satisfied.

\section{Scalar Field Inflationary Models and the Planck 2018 Constraints}

In the following we will examine for which values of the
dimensionless parameter $\alpha$ do some inflationary models
appearing in the Planck 2018 release of inflation
\cite{Akrami:2018odb}, comply with the recent Planck observational
data \cite{Akrami:2018odb}. The data indicate that the values of
the tensor-to-scalar ratio $r$ and the spectral index of the
primordial curvature perturbations $n_s$ are restricted as
follows,
\begin{equation}\label{PlankConstraints}
    n_s=0.9649 \pm 0.0042 \ , \ r<0.056.
\end{equation}

\subsection{D-Brane (p=2)}

\begin{figure}
\centering
\includegraphics[width=18pc]{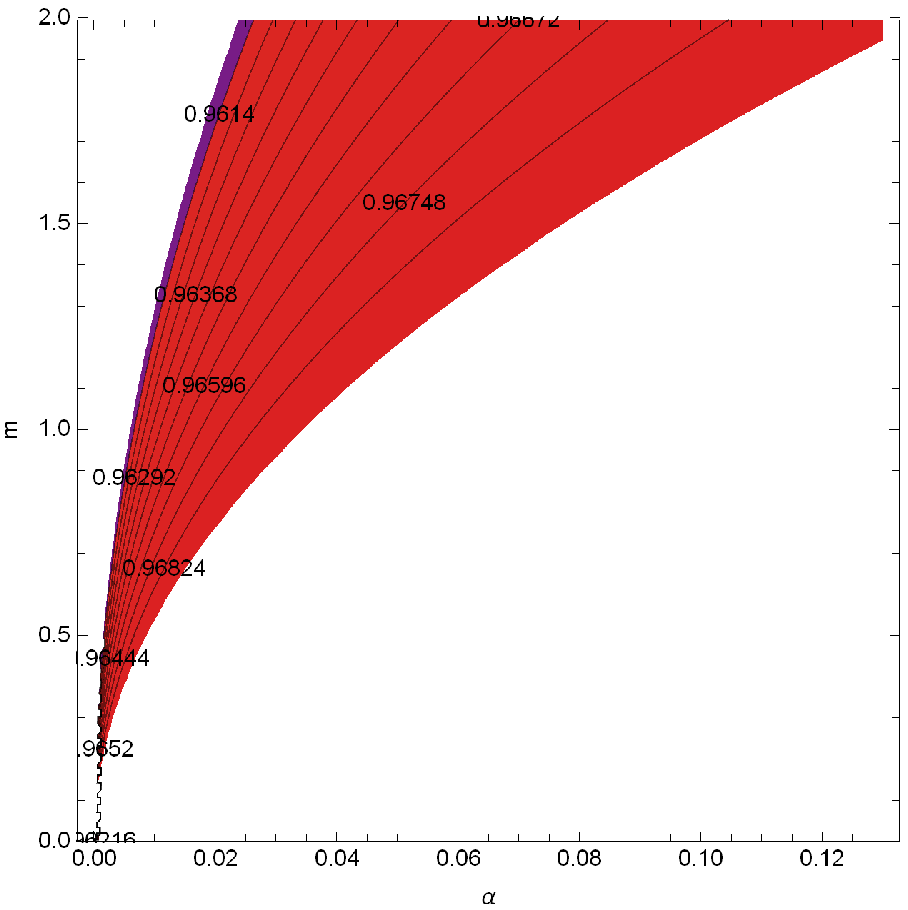}
\includegraphics[width=18pc]{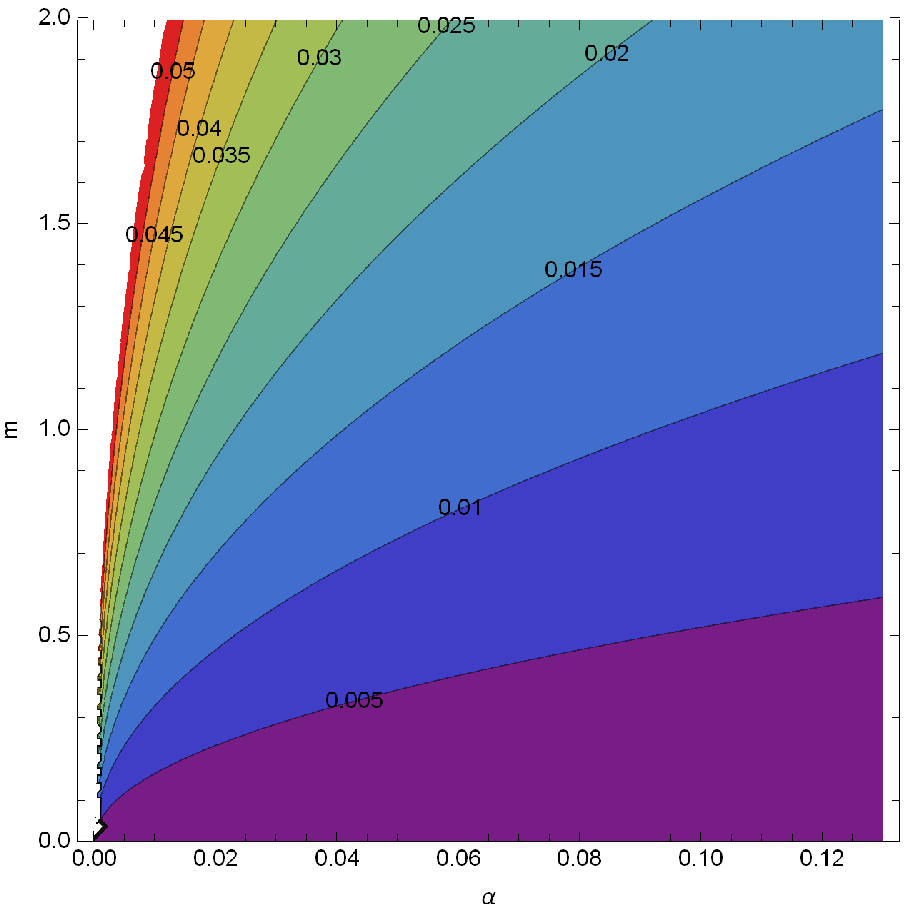}
\caption{Contour plot for the spectral index of primordial scalar
curvature perturbations $n_s$ (left plot) and the tensor-to-scalar
ratio $r$ (right plot) for $\alpha=[0, 0.13]$, $m=[10^{-6},
10^{0.3}]$ and $N=60$ for the D-Brane Model (p=2).}\label{Db2nspl}
\end{figure}

The first model to deal with is a D-Brane model
\cite{Akrami:2018odb},
\begin{equation}\label{D-Brane2}
V(\phi)=\Lambda ^4 \left(1-\left(\frac{m}{\kappa\phi}\right)^2\right),
\end{equation}
where $\Lambda$ has dimensions of mass [m] and $m$ is a
dimensionless parameter taking values in $[10^{-6}, 10^{0.3}]$.
Using equation (\ref{eps1 to eps}) we can obtain the first
slow-roll index for this model,
\begin{equation}\label{Db2e1}
    \epsilon_1 \simeq \frac{2 \alpha  m ^4}{\kappa^6 \phi ^6}.
    \end{equation}
By solving the equation $\epsilon_1(\phi_f)=1$ we find the value
of $\phi$ at the end of the inflationary era,
\begin{equation}\label{Db2ff}
    \phi_f=\frac{\sqrt[6]{2} \sqrt[6]{\alpha } m ^{2/3}}{\kappa},
\end{equation}
and using the integral in (\ref{N}) we get the value of $\phi$ at
the beginning of this era as well to be,
\begin{equation}\label{D2fi}
    \phi_i=\frac{\sqrt[4]{2^{2/3} \alpha ^{2/3} m ^{8/3}+8 \alpha  m ^2 N}}{\kappa}.
\end{equation}
Now, the expressions of the spectral index and the
tensor-to-scalar ratio at this point at leading order in $1/N$
are,
\begin{equation}\label{Db2ns}
    n_s \simeq \frac{3 m ^{2/3}}{8 \sqrt[3]{2} \sqrt[3]{\alpha } N^2}-\frac{3 \sqrt{\alpha  m ^2 N}}{4 \sqrt{2} \alpha  N^2}-\frac{3}{2 N}+1,
\end{equation}
\begin{equation}\label{Db2r}
   r \simeq \frac{\sqrt{2} \sqrt{\alpha  m ^2 N}}{\alpha  N^2}.
\end{equation}\\
Considering that both $n_s$ and $r$ should comply with the Planck
constraints in (\ref{PlankConstraints}) we should search for the
values of $\alpha$ and $m$ that satisfy these. Assuming that $N
\simeq 60$, we can construct the plots of Fig. \ref{Db2nspl} that
have been restricted to show only the regions that $n_s=0.9649 \pm
0.0042$ and $r<0.056$. Therefore, we conclude that the constraints
are met if,
\begin{equation}\label{Db2con}
    5.3933 \leq \frac{m}{\sqrt{\alpha}} \leq 12.9311.
\end{equation}
This can be written equivalently as,
 \begin{equation}\label{Db2cona}
     0.00598m^2 \leq \alpha \leq 0.03438m^2.
 \end{equation}
Generally, $\alpha=[0, 1]$ but since $m=[10^{-6}, 10^{0.3}]$ the
maximum value of $\alpha$ is $\alpha=0.1368$ when $m=10^{0.3}$ and
in this case $n_s=0.969099$ and $r=0.0164$.

\subsection{D-Brane (p=4)}

\begin{figure}
\centering
\includegraphics[width=18pc]{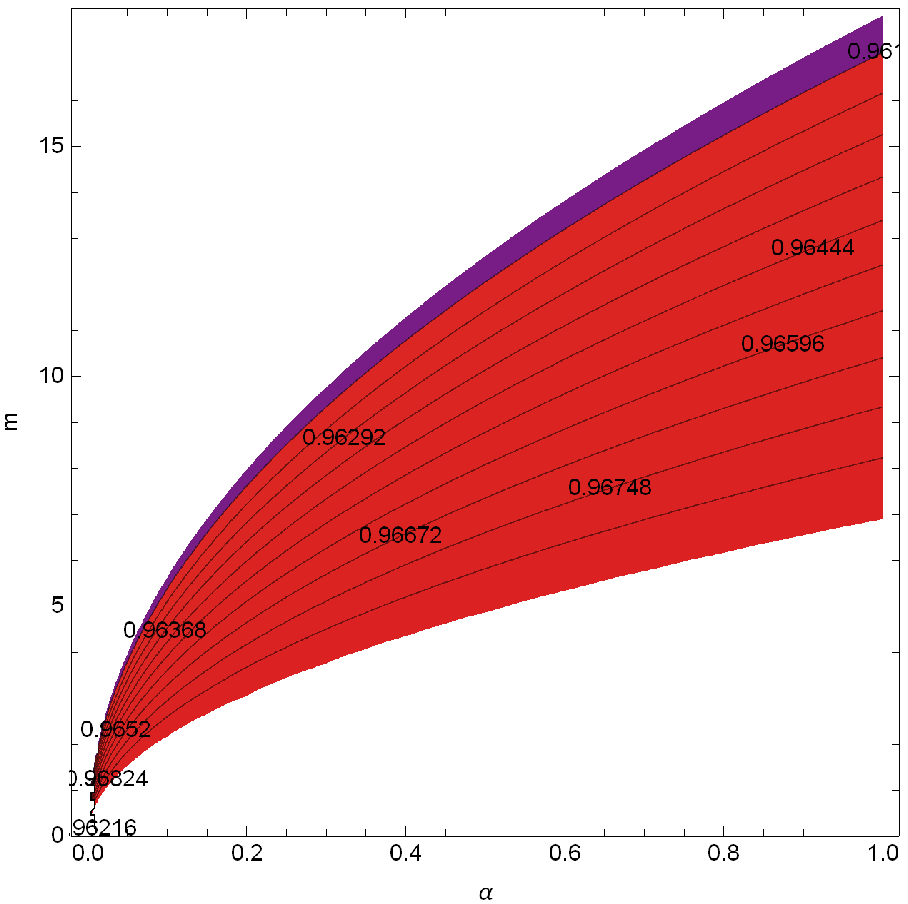}
\includegraphics[width=18pc]{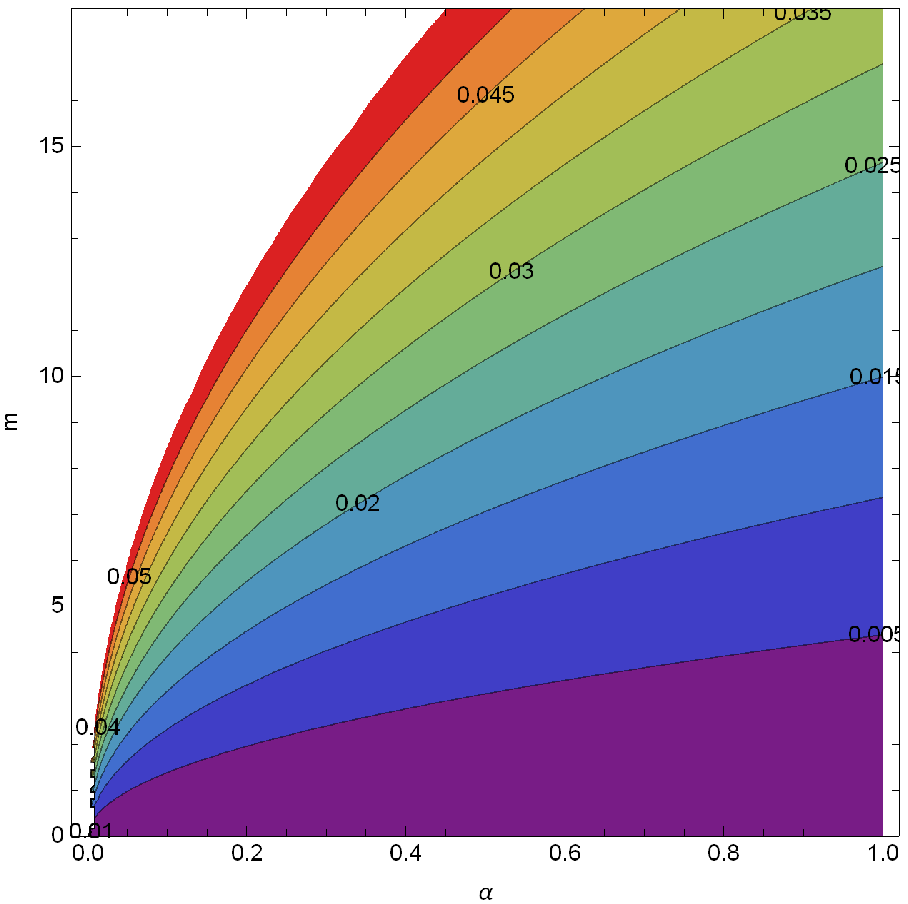}
\caption{Contour plot for the spectral index of primordial scalar curvature perturbations $n_s$ (left plot) and the tensor-to-scalar ratio $r$ (right plot) for $\alpha=[0, 1]$, $m=[10^{-6}, 18]$ and $N= 60$ for the D-Brane Model (p=4).}\label{Dbrane4nspl}
\end{figure}

We proceed to study another D-Brane inflation model,
\begin{equation}\label{DBrane4}
V(\phi)=\Lambda ^4 \left(1-\left(\frac{m}{\kappa\phi}\right)^4\right),
\end{equation}
where $\Lambda$ has dimensions of mass $[m]$ and $m$ is a
dimensionless parameter. For this model the first slow-roll index
is,
\begin{equation}\label{Dbrane4e1}
    \epsilon_1 \simeq \frac{8 \alpha  m ^8}{\kappa^{10} \phi
    ^{10}}.
    \end{equation}
We can obtain the value of $\phi$ at the end of inflation by
solving the equation $\epsilon_1 (\phi_f)=1$ and, thus, we get
that,
\begin{equation}\label{Dbrane4ff}
    \phi_{f}=\frac{2^{3/10} \sqrt[10]{\alpha } m ^{4/5}}{\kappa}.
\end{equation}
From the integral of (\ref{N}) solved with respect to $\phi_i$ we find,
\begin{equation}\label{Dbrane4fi}
    \phi_i=\frac{\sqrt[6]{2} \sqrt[6]{2^{4/5} \alpha ^{3/5} m ^{24/5}+12 \alpha  m ^4 N}}{\kappa},
\end{equation}
and we may obtain the expressions of the tensor-to-scalar ratio
$r$ and the spectral index $n_s$ at the beginning of the
inflationary era at leading order in $1/N$,
\begin{equation}\label{DBranens}
    n_s \simeq \frac{5 m ^{4/5}}{18 \sqrt[5]{2} \alpha ^{2/5} N^2}-\frac{\sqrt[3]{\alpha  m ^4 N}}{2\ 3^{2/3} \alpha  N^2}-\frac{5}{3 N}+1,
\end{equation}
\begin{equation}\label{Dbraner}
   r \simeq \frac{4 \sqrt[3]{\alpha  m ^4 N}}{3\ 3^{2/3} \alpha  N^2}.
\end{equation}
There is a variety of values for $\alpha$ and $m$ such that the
constraints on $n_s$ and $r$ dictated in (\ref{PlankConstraints})
are met for $N \simeq 60$, as it can be seen from the plots in
Fig. \ref{Dbrane4nspl} that have a range restricted in accordance
with the Planck constraints. For extended precision we can solve
the respective algebraic inequalities and can easily be found that
$n_s$ and $r$ are simultaneously compatible with the Planck data
for every pair of $\alpha$ and $m$ that satisfy
\begin{equation}\label{Dbrane4con1}
    0.00313782 \leq \frac{\alpha}{m^2} \leq 0.0209707.
\end{equation}
Recalling that, generally, $\alpha=[0, 1]$, an alternative form of
the range of values of $\alpha$ and $m$ is,
\begin{align}\label{T2mcon}
   & 0.00313782 m^2 \leq \alpha \leq 0.0209707m^2 \ , \ m=[10^{-6}, 6.906] \\ \notag &
  0.00313782 m^2 \leq \alpha \leq 1 \ , \ m=[6.906, 17.852].
\end{align}
Let us give a numerical example. For instance, in the case that
$N=60$ and $m=5$, then $\alpha=[0.078, 0.524]$ and say we have
$\alpha=0.2$ then $n_s=0.9662$ and $r=0.0174$, which are in
accordance with the Planck data.

\subsection{Natural Inflation}

\begin{figure}
\centering
\includegraphics[width=18pc]{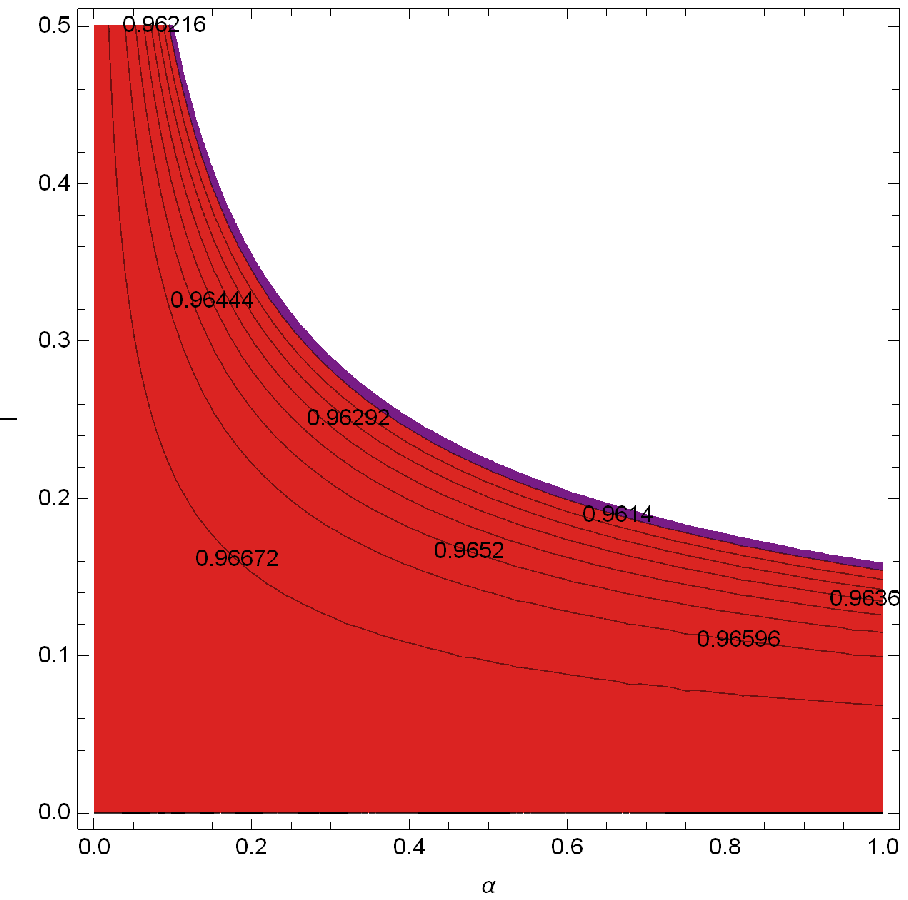}
\includegraphics[width=18pc]{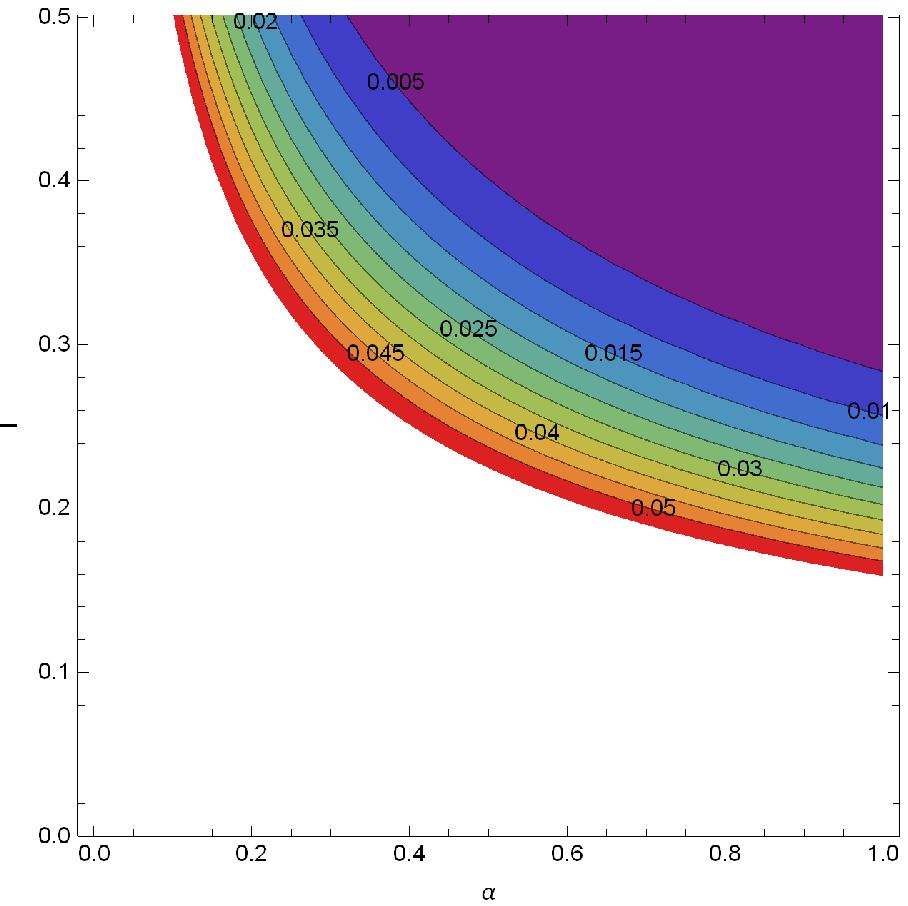}
\caption{Contour plot for the spectral index of primordial scalar
curvature perturbations $n_s$ (left plot) and the tensor-to-scalar
ratio $r$ (right plot) for $\alpha=[0,1]$, $l=[0.0032,0.50112]$
and $N= 60$ for the Natural inflation model.}\label{Natnspl}
\end{figure}

The Natural inflation model is described by the potential,
\begin{equation} \label{natural_potential}
V(\phi)=\Lambda^4(1-\cos{(l\kappa\phi)}),
\end{equation}
where $\Lambda$ has dimensions of mass $[m]$ and $l$ is a
dimensionless constant that generally takes values in $[10^{0.3} ,
10^{2.5}]$. For this potential, the first slow-roll index takes
the form,
\begin{equation}\label{Nate1}
    \epsilon_1=\frac{1}{2} \alpha  l ^2 \tan^2{\left(\frac{\kappa l \phi }{2}\right)}.
\end{equation}
We obtain $\phi$ at the end of the inflationary era from the
equation $\epsilon_1(\phi_f)=1$ and we get,
\begin{equation}\label{Natff}
    \phi_f=\frac{2 \tan ^{-1}{\left(\frac{\sqrt{2}}{\sqrt{\alpha } l }\right)}}{\kappa l }.
\end{equation}
Proceeding, we use the integral that gives the number of
$e$-foldings $N$ from (\ref{N}) to find $\phi_i$, which is,
\begin{equation}\label{Natfi}
  \phi_i=\frac{2 \sin ^{-1}{ \bigg\{ e^{ \bigg[\frac{1}{2} \left(2 \log \left(\frac{\sqrt{2}}{\sqrt{\alpha } l  \sqrt{\frac{2}{\alpha  l ^2}+1}}\right)-\alpha  l ^2 N\right)\bigg]}\bigg\}}}{\kappa l }.
\end{equation}
Our interest lies in the spectral index $n_s$ and the
tensor-to-scalar ratio $r$ so, for this model at $\phi_i$ their
expressions are,
\begin{equation}\label{Natns}
    n_s=-\frac{2 \alpha  l ^2+\left(\alpha ^2l ^4+\alpha l ^2-2\right) e^{\alpha l ^2 N}+2}{\left(\alpha l ^2+2\right) e^{\alpha l ^2 N}-2},
\end{equation}
\begin{equation}\label{Natr}
    r=\frac{16 \alpha l ^2}{\left(\alpha l ^2+2\right) e^{\alpha l ^2 N}-2}.
\end{equation}
We are looking for the values of $\alpha$ and $l$ that allow $n_s$
and $r$ to lie within the range of the Planck constraints in
(\ref{PlankConstraints}). As can be seen from the plots in Fig.
\ref{Natnspl} that depict the behavior of $n_s$ and $r$ for $N
\simeq 60$ and have restricted in the constraints regions, but
also solving the corresponding inequalities, we find that the
constraints are met if,
\begin{equation}\label{Natcon}
    0.02525<\alpha l^2 <0.02553,
\end{equation}
or with respect to $\alpha$
\begin{equation}\label{Natcona}
    \frac{0.02525}{l^2}<\alpha<\frac{0.02553}{l^2}.
\end{equation}
We observe that the region of values for the parameter $\alpha$
for each $l$ that renders this potential viable is very narrow.
Let us provide a numerical example for further clarification.
Supposing that $l=10$, the permitted values of $\alpha$ that make
this potential agree with the Planck constraints are
$\alpha=[2.525 \times 10^{-4}, 2.553 \times 10^{-4}]$ and we
select $\alpha=2.55 \times 10^{-4}$, a really small value, then
$n_s=0.96073$ and $r=0.0559$, in the borderline of the
constraints.

\subsection{T-Model (m=1)}

\begin{figure}
\centering
\includegraphics[width=18pc]{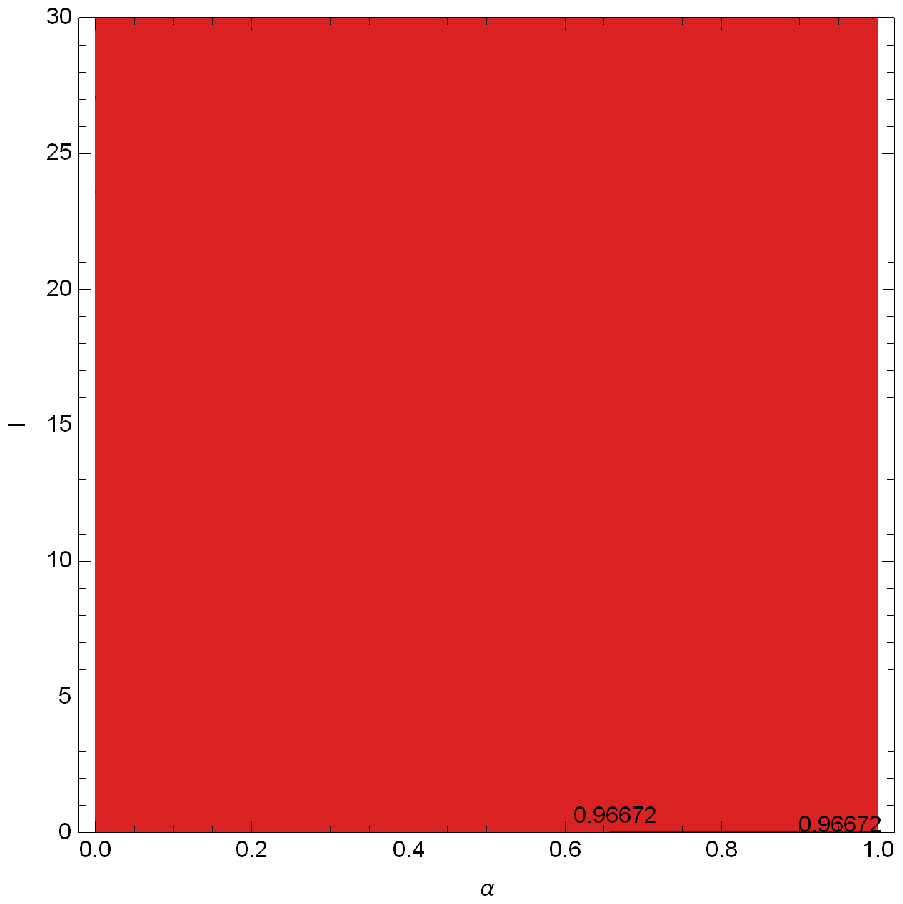}
\includegraphics[width=18pc]{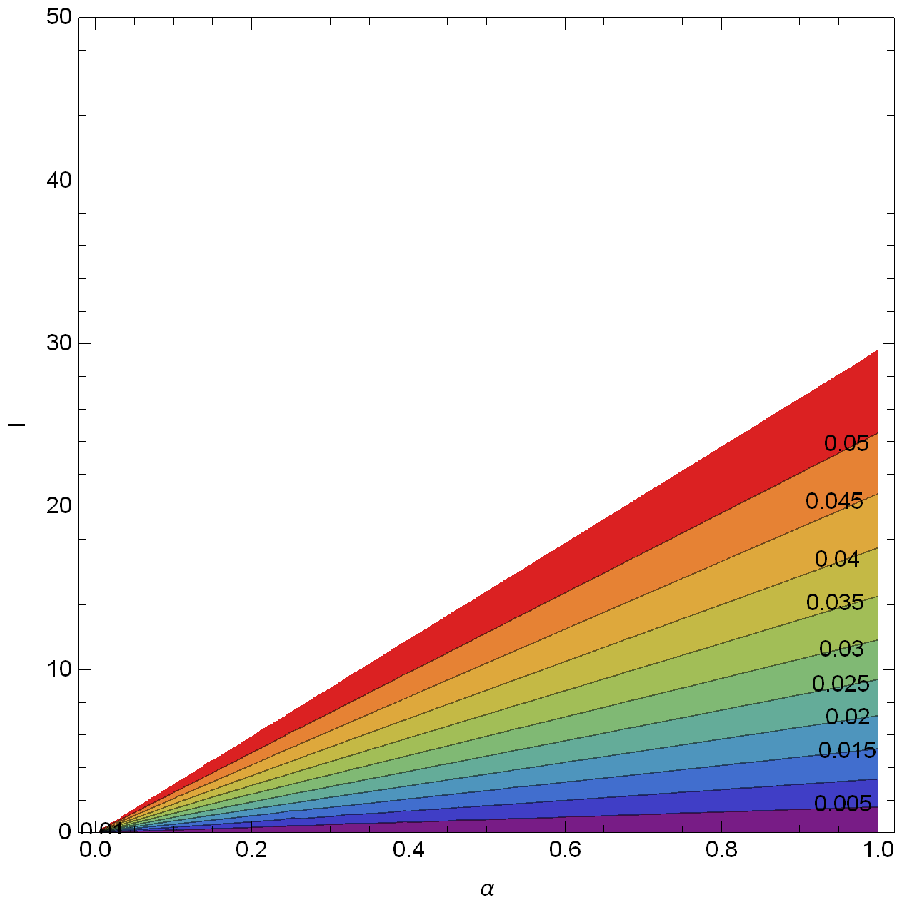}
\caption{Contour plot for the spectral index of primordial scalar
curvature perturbations $n_s$ for $\alpha=[0, 1]$, $l=[10^{-2},
30]$ (left plot) and the tensor-to-scalar ratio $r$ (right plot)
for $\alpha=[0, 1]$, $l=[10^{-2}, 50]$ and $N= 60$ for the T-Model
(m=1).}\label{T1nspl}
\end{figure}

The next inflationary model to study belongs to the family of the
T-Models and is described by the following potential,
\begin{equation}\label{T1V}
V(\phi)=\Lambda ^4 \tanh^2{\left(\frac{\kappa \phi}{\sqrt{6 l}}\right)},
\end{equation}
where, as usual, $\Lambda$ has dimensions of mass $[m]$ and $l$ is
a dimensionless parameter with values within the range $[10^{-2},
10^{4}]$. The first slow-roll index in this case is,
\begin{equation}
\centering\label{T1e1}
    \epsilon_1 = \frac{4 \alpha\, \mathrm{csch} ^2 \left(\frac{\sqrt{\frac{2}{3}} \kappa \phi }{\sqrt{l }}\right)}{3 l }.
\end{equation}
Once more, we solve the equation $\epsilon_1(\phi_f)=1$ and we obtain $\phi$ at the end of inflation,
\begin{equation}\label{T1ff}
    \phi_f=\frac{\sqrt{\frac{3}{2}} \sqrt{l } \sinh ^{-1}{\left(\frac{2 \sqrt{\alpha }}{\sqrt{3} \sqrt{l }}\right)}}{\kappa},
\end{equation}
and from the integral of (\ref{N}) we find the initial $\phi_i$,
\begin{equation}\label{T1fi}
    \phi_i=\frac{\sqrt{\frac{3}{2}} \sqrt{l } \cosh ^{-1}{\left(\frac{3 l  \sqrt{\frac{4 \alpha }{3 l }+1}+4 \alpha  N}{3 l }\right)}}{\kappa},
\end{equation}
where $N$ is the number of $e$-foldings the Universe has grown
during the inflationary era. The spectral index of the primordial
curvature perturbations $n_s$ and the tensor-to-scalar ratio $r$
for this potential are,
\begin{align}\label{T1ns}
    n_s=&-\frac{18 l }{3 l +4 \alpha  N^2+2 l  N \sqrt{\frac{12 \alpha }{l }+9}}+\\ \notag &\frac{2 \alpha  \left(\text{csch}^2\left(\frac{1}{2} \cosh ^{-1}\left(\frac{l  \sqrt{\frac{12 \alpha }{l }+9}+4 \alpha
   N}{3 l }\right)\right)-2\right) \text{sech}^2\left(\frac{1}{2} \cosh ^{-1}\left(\frac{l  \sqrt{\frac{12 \alpha }{l }+9}+4 \alpha  N}{3 l }\right)\right)}{3 l }+1,
\end{align}
\begin{equation}\label{T1r}
   r=\frac{48 l }{3 l +4 \alpha  N^2+2 l  N \sqrt{\frac{12 \alpha }{l }+9}}.
\end{equation}
This model can provide a good inflationary phenomenology for a
variety of $\alpha$ and $l$ values. Specifically, $n_s$ and $r$
comply with the Planck constraints when $N \simeq 60$ for,
\begin{equation}\label{T1con}
    \frac{\alpha}{l} > 0.03376,
\end{equation}
or equivalently, considering that generally $\alpha=[0, 1]$, for,
\begin{equation}\label{T1cona}
    0.03376 l < \alpha \leq 1.
\end{equation}
Of course, this demands that $l \leq 29.62$ and combined with the
original restrictions on $l$ we get that $l=[10^{-2}, 29.62]$. A
better visualization of these constraints can be given by the
plots in Fig. \ref{T1nspl} which have been limited in containing
the regions that satisfy the constraints on $n_s$ and $r$ for $N
\simeq 60$. A numerical example should simplify things thus, let
us consider that $N=60$, $l=2.5$ then $\alpha=[0.844, 1]$ and we
select $\alpha=0.88$, then $n_s=0.9669$ and $r=0.0087$, values
that are in accordance with the Planck constraints.

\subsection{T-Model (m=2)}

\begin{figure}
\centering
\includegraphics[width=18pc]{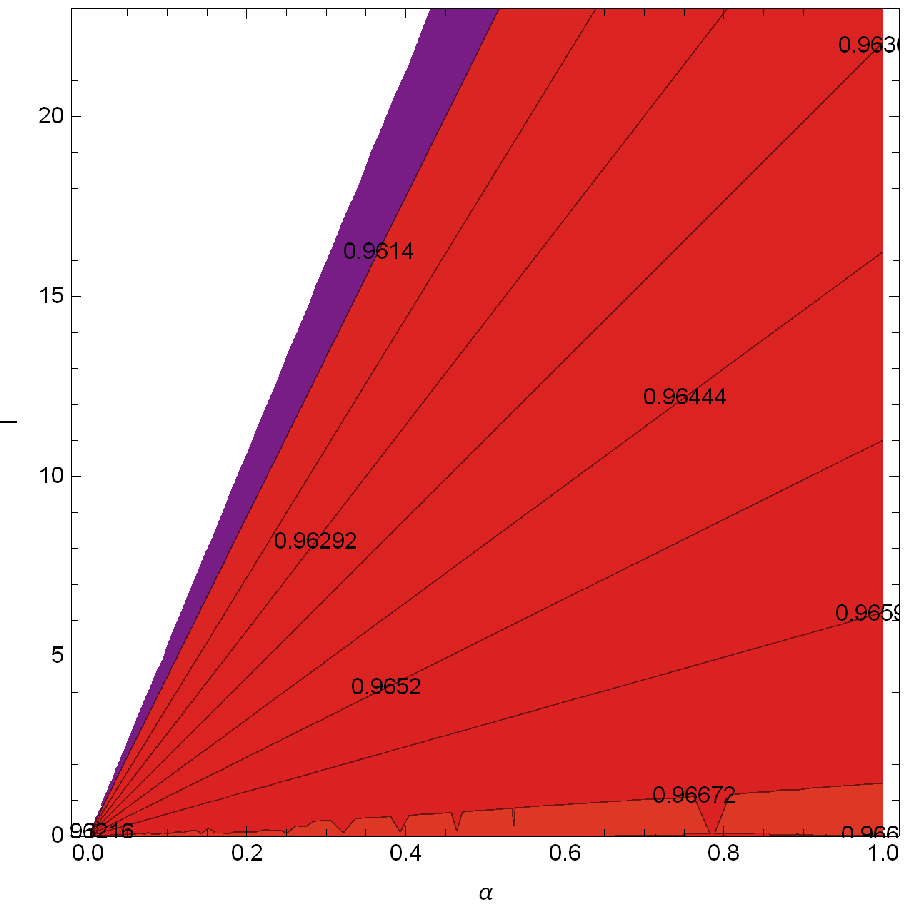}
\includegraphics[width=18pc]{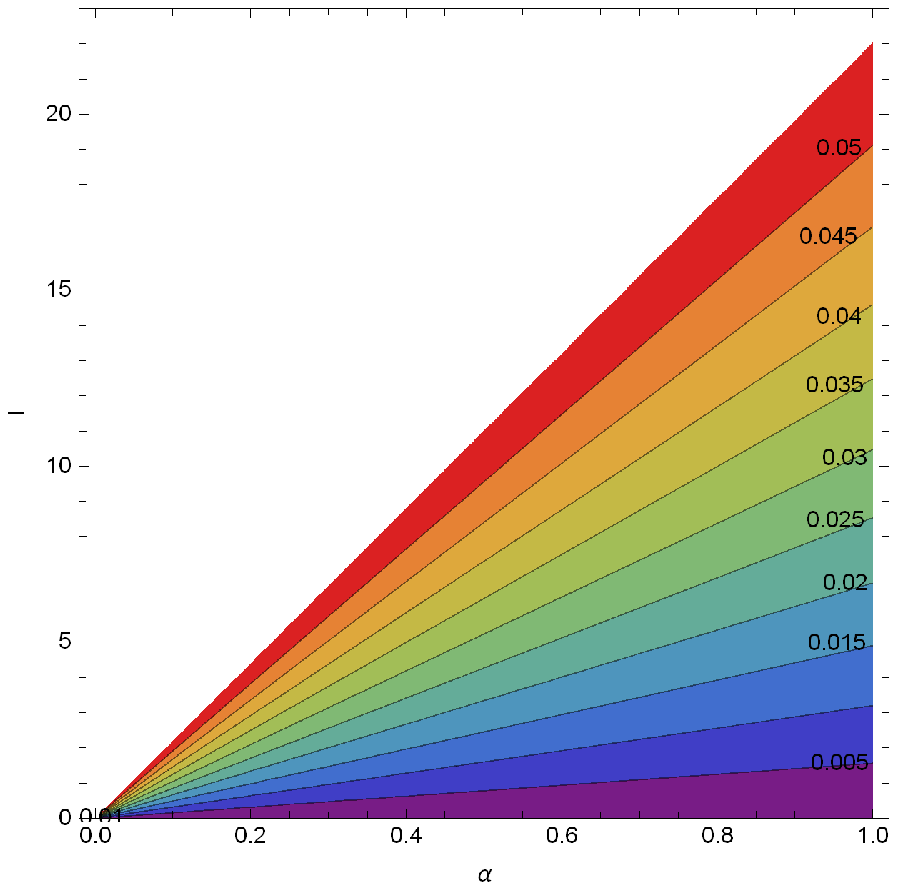}
\caption{Contour plot for the spectral index of primordial scalar curvature perturbations $n_s$ (left plot) and the tensor-to-scalar ratio $r$ (right plot) for $\alpha=[0, 1]$, $l=[0.01, 23]$ and $N= 60$ for the T-Model (m=2).}\label{T2nspl}
\end{figure}

Another T-Model that is interesting to study is described by the following potential,
\begin{equation}\label{T1}
V(\phi)=\Lambda ^4 \tanh^4{\left(\frac{\kappa \phi}{\sqrt{6 l}}\right)},
\end{equation}
with the dimensions of $\Lambda$ being once again these of mass $[m]$ and $l$ is dimensionless and generally takes values in $[10^{-2}, 10^{4}]$.\\
This potential's first slow-roll index is,
\begin{equation}\label{T2e1}
    \epsilon_1=\frac{16 \alpha  \, \mathrm{csch}^2 {\left(\frac{\sqrt{\frac{2}{3}} \kappa \phi }{\sqrt{l }}\right)}}{3
   l }.
\end{equation}
By setting $\epsilon_1=1$ we obtain $\phi_f$, $\phi$ at the end of
inflation, which is,
\begin{equation}\label{T2ff}
    \phi_f=\frac{\sqrt{\frac{3}{2}} \sqrt{l } \sinh ^{-1}\left(\frac{4 \sqrt{\alpha }}{\sqrt{3}
   \sqrt{l }}\right)}{\kappa}.
\end{equation}
The next step is to derive $\phi_i$ from (\ref{N}) which gives,
\begin{equation}\label{T2fi}
    \phi_i=\frac{\sqrt{\frac{3}{2}} \sqrt{l } \cosh ^{-1}\left(\frac{3 l \sqrt{\frac{16 \alpha }{3
   l }+1}+8 \alpha  N}{3 l }\right)}{\kappa}.
\end{equation}
Therefore, the spectral index and the tensor-to-scalar ratio for
this model are,
\begin{align}\label{T2ns}
    n_s=&-\frac{18 l }{3 l +4 \alpha  N^2+l  N \sqrt{\frac{48 \alpha }{l }+9}}+\\ \notag & \frac{4 \alpha
    \left(3 \text{csch}^2\left(\frac{1}{2} \cosh ^{-1}\left(\frac{l  \sqrt{\frac{48 \alpha
   }{l }+9}+8 \alpha  N}{3 l }\right)\right)-2\right) \text{sech}^2\left(\frac{1}{2}
   \cosh ^{-1}\left(\frac{l  \sqrt{\frac{48 \alpha }{l }+9}+8 \alpha  N}{3 l
   }\right)\right)}{3 l }+1,
\end{align}
\begin{equation}\label{T2r}
   r=\frac{48 l }{3 l +4 \alpha  N^2+l  N \sqrt{\frac{48 \alpha }{l }+9}}.
\end{equation}
This model also provides with a good phenomenology regarding
inflation for a variety of values of $\alpha$ and $l$. As can be
observed from the plots in Fig. \ref{T2nspl} that are bounded
according to the Planck constraints to depict the regions that
$n_s=0.9649 \pm 0.0045$ and $r<0.056$ for $N \simeq 60$, but as
can also be found algebraically for the same $N$, in order to meet
these constraints we need
\begin{equation}\label{T2con}
    0.04538 < \frac{\alpha}{l} <30.
\end{equation}
However, considering that the initial ranges of $\alpha$ and $l$
are $[0, 1]$ and $[10^{-2}, 10^{4}]$ respectively, the above
condition poses further restrictions on $l$ and can be rewritten
as,
\begin{align}\label{T2mcon}
   & 0.04538l \leq \alpha \leq 30l \ , \ l=[10^{-2},0.03] \\ \notag &
   0.0453846l \leq \alpha \leq 1 \ , \ l=[ 0.033,22.034].
\end{align}
Let us consider the values $ N=60$, $l= 0.5$ and $\alpha= 0.04$,
and we get $n_s=0.9649$ and $r=0.035$, values that satisfy the
Planck constraints.

\subsection{Potential with Exponential Tails Model}

\begin{figure}
\centering
\includegraphics[width=18pc]{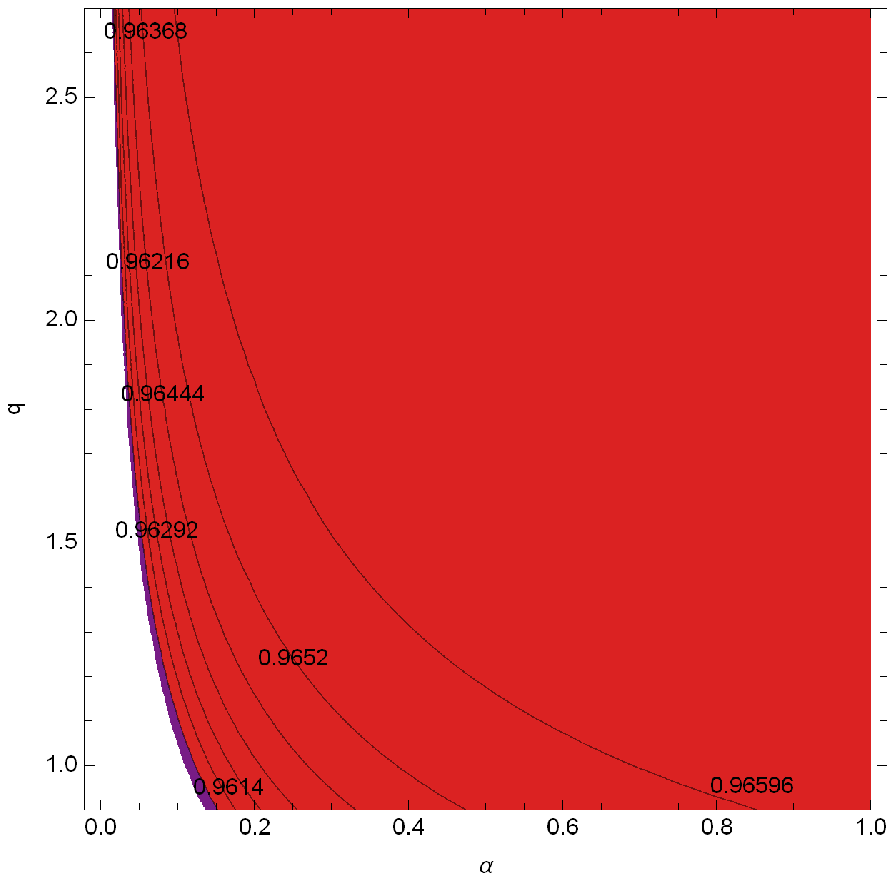}
\includegraphics[width=18pc]{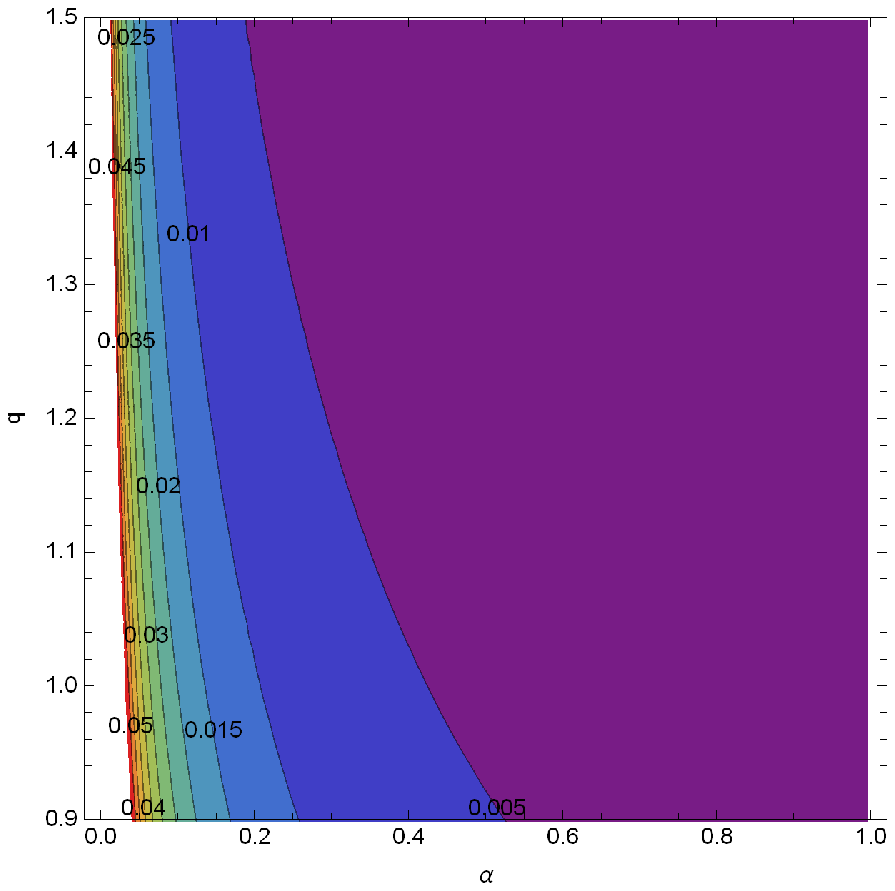}
\caption{Contour plot for the spectral index of primordial scalar
curvature perturbations $n_s$ (left plot) and the tensor-to-scalar
ratio $r$ for $\alpha=[0, 1]$, $q=[0.9, 2.7]$ (right plot) for
$\alpha=[0, 1]$, $q=[0.9, 1.5]$ and $N=60$ for the inflation with
exponential tails model.}\label{ETnspl}
\end{figure}

The next model to be tested when it comes to its inflationary
phenomenology is the potential with exponential tails model with
the potential,
\begin{equation}\label{ETV}
    V(\phi)=\Lambda^4(1-e^{-\kappa q \phi }),
\end{equation}
where $\Lambda$ has dimensions of mass $[m]$ and $q$ is a
dimensionless parameter within the range $[10^{-3}, 10^{3}]$. The
first slow-roll index for this model is,
\begin{equation}\label{ETe1}
    \epsilon_1=\frac{\alpha  q^2}{2 \left(e^{\kappa q \phi }-1\right)^2},
\end{equation}
and from $\epsilon_1(\phi_f)=1$ we obtain the expression of $\phi$
at the end of the inflationary era, which is,
\begin{equation}\label{ETff}
    \phi_f=\frac{\log \left(\frac{1}{2} \left(\sqrt{2} \sqrt{\alpha } q+2\right)\right)}{\kappa q}.
\end{equation}
Now, we proceed to finding the expression of $\phi$ via the integral in (\ref{N}) and we get,
\begin{equation}\label{ETfi}
    \phi_i \simeq \frac{\log \left(\frac{1}{2} \left(2 \alpha  q^2 N+\sqrt{2} \sqrt{\alpha } q+2\right)\right)}{\kappa q}.
\end{equation}
Hence, this model's spectral index and tensor-to-scalar ratio are,
respectively,
\begin{equation}\label{ETns}
    n_s=-\frac{4 \alpha  q}{\sqrt{2} \sqrt{\alpha }+2 \alpha  q N}-\frac{12}{\left(2 \sqrt{\alpha } q N+\sqrt{2}\right)^2}+1,
\end{equation}
\begin{equation}\label{ETr}
    r=\frac{32}{\left(2 \sqrt{\alpha } q N+\sqrt{2}\right)^2}.
\end{equation}
We are interested in the values of $\alpha$ and $q$ for which
$n_s$ and $r$ comply with the Planck constraints. As can be seen
from the plots in Fig. \ref{ETnspl} that are calculated for $N
\simeq 60$ and have been bounded to show the regions that $n_s$
and $r$ satisfy the aforementioned constraints, there is a variety
of $\alpha$, $l$ pairs of values that do so. Strictly
mathematically, for $N \simeq 60$ the restrictions on these
parameters, taking into account their initial ranges, are,
\begin{align}\label{T2mcon}
   & 0.993/q^2 \leq \alpha \leq 1\ , \ q=[0.9967, 1.0059] \\ \notag &
   0.993/q^2 \leq \alpha \leq 1.012/q^2  \ , \ q=[1.0059, 10^3].
\end{align}

\subsection{E-Model (n=1)}

\begin{figure}
\centering
\includegraphics[width=18pc]{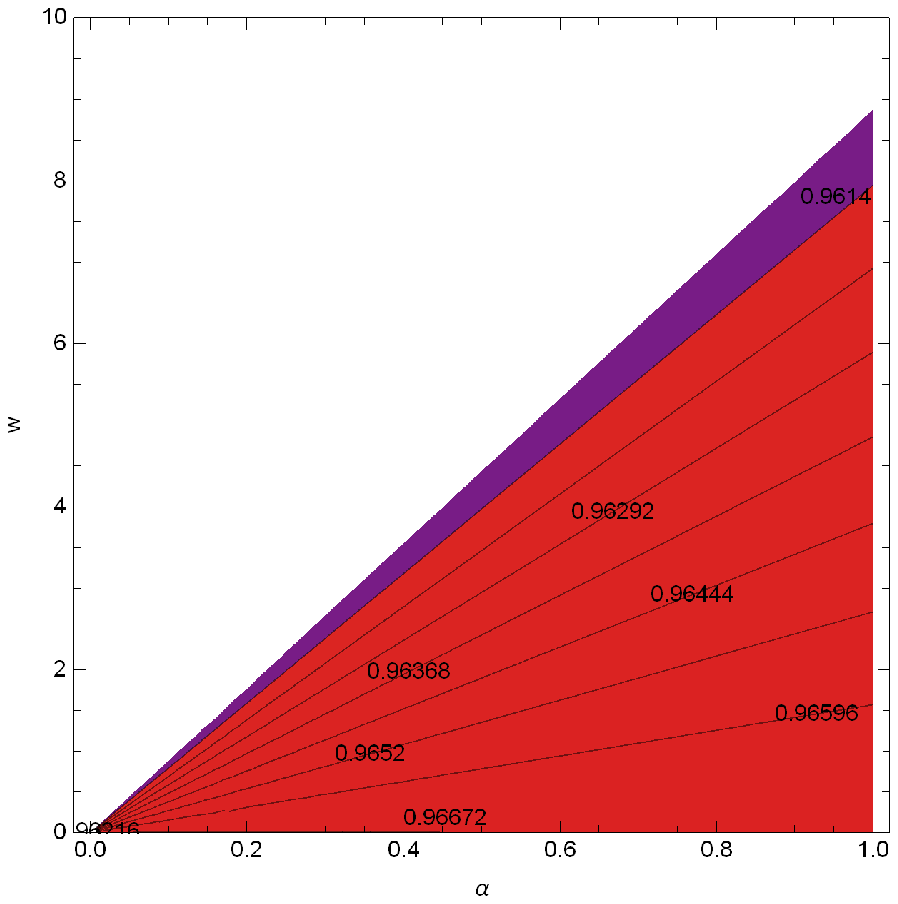}
\includegraphics[width=18pc]{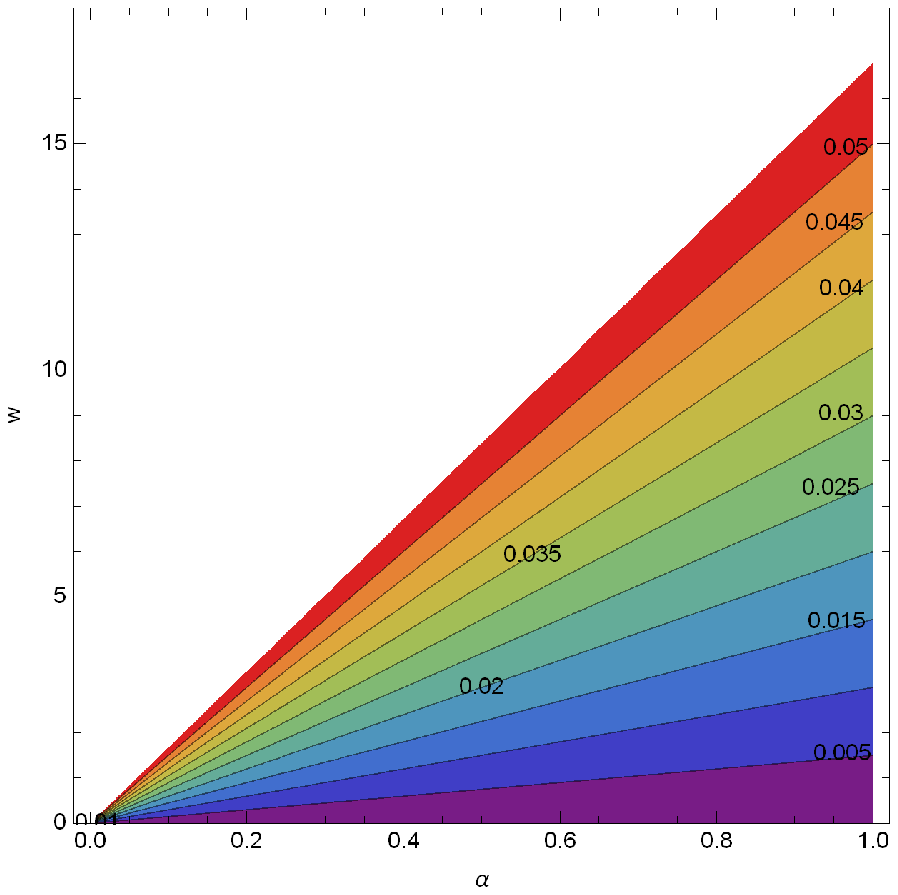}
\caption{Contour plot for the spectral index of primordial scalar
curvature perturbations $n_s$ for $\alpha=[0, 1]$, $w=[0, 10]$
(left plot) and the tensor-to-scalar ratio $r$ (right plot) for
$\alpha=[0, 1]$, $w=[0, 18]$ and $N=60$ for the E-Model
(n=1).}\label{E1nspl}
\end{figure}

Another interesting family of Inflationary models are the E-Models. One of them is described by the potential,
\begin{equation}\label{E1}
V(\phi)=\Lambda ^4 \left(1-e^{-\frac{\sqrt{\frac{2}{3}} \kappa \phi}{\sqrt{w}}}\right)^2 ,
\end{equation}
where $\Lambda$ has dimensions of mass $[m]$ and $w$ is a
dimensionless parameter that takes values within the range
$[10^{-2}, 10^4]$. The expression of the first slow-roll parameter
for this potential is,
\begin{equation}\label{E1e1}
    \epsilon_1=\frac{4 \alpha }{3 w  \left(e^{\sqrt{\frac{2}{3}} \kappa \sqrt{\frac{1}{w }} \phi }-1\right)^2}.
\end{equation}
To obtain $\phi$ at the end of inflation we have to solve the
equation $\epsilon_1(\phi_f)=1$ and we find that,
\begin{equation}\label{E1ff}
    \phi_f=\frac{\sqrt{\frac{3}{2}} \log \left(\frac{2 \sqrt{3} \sqrt{\alpha }+3 \sqrt{w }}{3 \sqrt{w }}\right)}{\kappa \sqrt{\frac{1}{w }}},
\end{equation}
and now using (\ref{N}) and solving with respect to $\phi_i$ we get,
\begin{equation}\label{E1fi}
    \phi_i=\frac{\sqrt{\frac{3}{2}} \log \left(\frac{2 \sqrt{3} \sqrt{\alpha } \sqrt{w }+3 w +4 \alpha  N}{3 w }\right)}{\kappa \sqrt{\frac{1}{w }}}.
\end{equation}
The spectral index of the primordial curvature perturbations and
the tensor-to-scalar ratio at the first horizon crossing at
leading order in $1/N$ are,
\begin{equation}\label{E1ns}
  n_s \simeq \frac{\sqrt{3} \sqrt{\alpha } \sqrt{w }-3 w +\alpha  N^2-2 \alpha  N}{\alpha  N^2},
\end{equation}
\begin{equation}\label{E1r}
  r \simeq \frac{12 w }{\alpha  N^2}.
\end{equation}
There are plenty of values of the parameters $\alpha$ and $w$ for
$N \simeq 60$ such that $n_s$ and $r$ comply with the Planck data
constraints (\ref{PlankConstraints}). One proof is given by the
plots in Fig. \ref{E1nspl} that have been limited to depict the
values of the spectral index and the tensor-to-scalar ratio that
satisfy the Planck constraints. Another proof can be given
mathematically, namely, solving the corresponding inequalities.
For both constraints to be met at the same time we need
\begin{equation}\label{E1con}
    0.308075 \leq \sqrt{\frac{\alpha}{w}}.
\end{equation}
Taking into consideration that generally $w=[10^{-2}, 10^4]$ and
$\alpha$ should belong within the range $[0, 1]$, the constraint
(\ref{E1con}) restricts the values of $\alpha$ and $w$ for $N
\simeq 60$ in,
\begin{equation}
    0.0949w \leq \alpha \leq 1 \ , \ w=[10^{-2}, 10.538].
\end{equation}
A numerical example would be useful. We suppose that $N=60$,
$w=5.2$ and $\alpha=0.6$, then $n_s=0.9609$ and $r=0.0289$, values
that satisfy the Planck conditions of (\ref{PlankConstraints}).

\subsection{E-Model (n=2)}

\begin{figure}
\centering
\includegraphics[width=18pc]{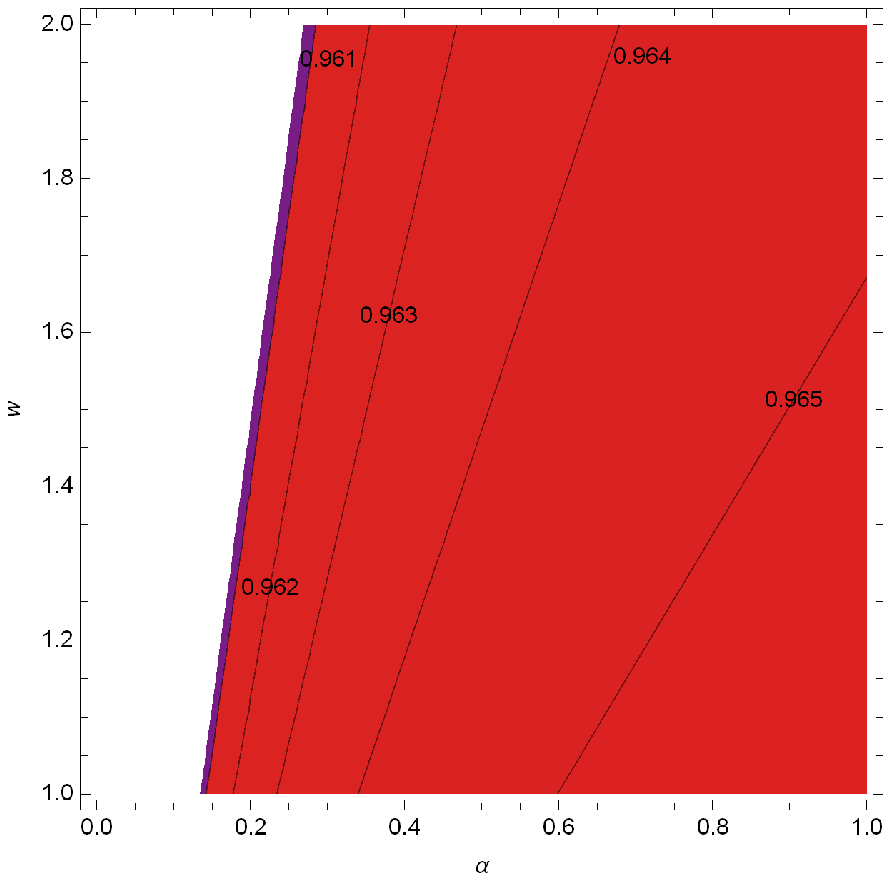}
\includegraphics[width=18pc]{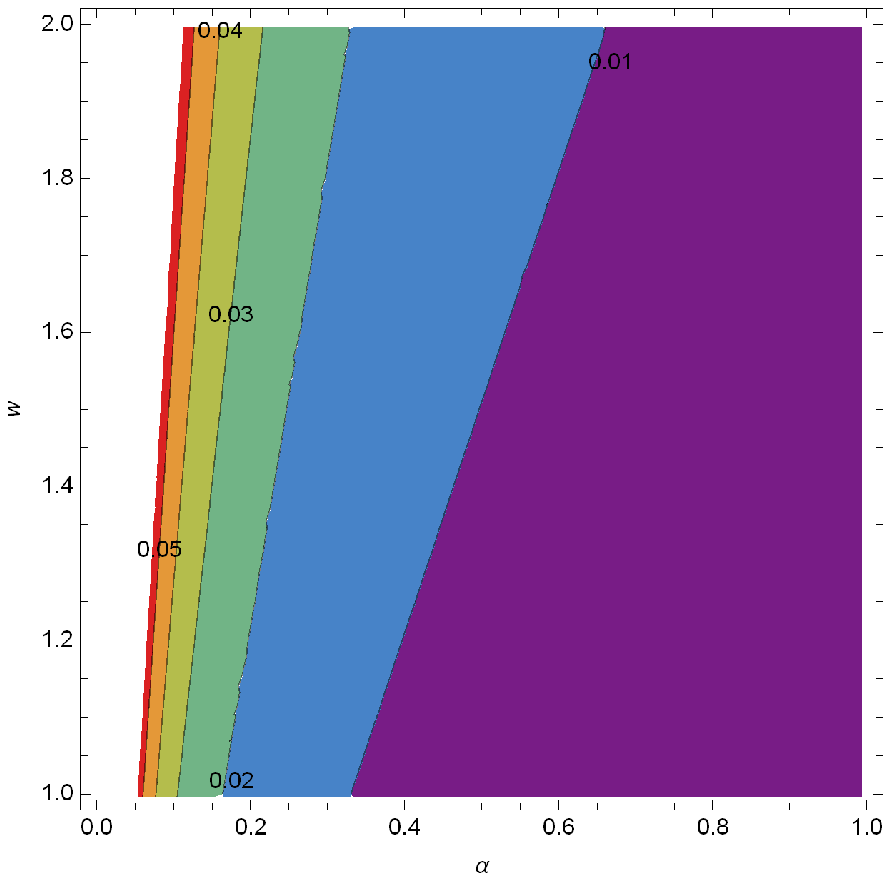}
\caption{Contour plot for the spectral index of primordial scalar
curvature perturbations $n_s$ (left plot) and the tensor-to-scalar
ratio $r$ (right plot) for $\alpha=[0, 1]$, $w=[0.9, 2]$ and
$N=60$ for the E-Model (n=2).}\label{E2nspl}
\end{figure}

We will study the inflationary phenomenology of yet another
E-Model with the potential,
\begin{equation}\label{E2}
V(\phi)=\Lambda ^4 \left(1-e^{-\frac{\sqrt{\frac{2}{3}} \kappa \phi}{\sqrt{w}}}\right)^4 .
\end{equation}
For this potential, the first slow-roll index is,
\begin{equation}\label{E2e1}
    \epsilon_1=\frac{16 \alpha }{3 w  \left(e^{\sqrt{\frac{2}{3}} \kappa \sqrt{\frac{1}{w }} \phi }-1\right)^2},
    \end{equation}
and by solving the equation $\epsilon_1(\phi_f)=1$ the expression
of $\phi_f$ can be found equal to,
\begin{equation}\label{E2ff}
    \phi_{f}=\frac{\sqrt{\frac{3}{2}} \log \left(\frac{4 \sqrt{\alpha }}{\sqrt{3} \sqrt{w }}+1\right)}{\kappa \sqrt{\frac{1}{w }}}.
\end{equation}
From (\ref{N}) we can find $\phi_i$,
\begin{equation}\label{fi}
    \phi_i=\frac{\sqrt{\frac{3}{2}} \log \left(\frac{4 \sqrt{\alpha }}{\sqrt{3} \sqrt{w }}+\frac{8 \alpha  N}{3 w }+1\right)}{\kappa \sqrt{\frac{1}{w }}}.
\end{equation}
Therefore, the spectral index $n_s$ and the tensor-to-scalar ratio
$r$ for this model are,
\begin{equation}\label{E2ns}
    n_s \simeq \frac{\frac{\sqrt{3} \sqrt{w }}{\sqrt{\alpha }}-\frac{9 w }{4 \alpha }}{N^2}-\frac{2}{N}+1 ,
\end{equation}
\begin{equation}\label{E2r}
   r \simeq \frac{12 w }{\alpha  N^2}.
\end{equation}
We are interested in finding which values of the parameters
$\alpha$ and $w$ render this model capable of describing the
inflationary era based on the Planck constraints in
(\ref{PlankConstraints}). Supposing that $N \simeq 60$ and by
solving the inequalities $n_s=0.9649 \pm 0.0042$ and $r<0.056$, we
conclude that we need,
\begin{equation}\label{E2con}
    0< \sqrt{\frac{w}{\alpha}}\leq 3.4986  .
\end{equation}
But $\alpha$ and $w$ should belong respectively in a subset of
$[0, 1]$ and [$10^{-2}, 10^4$], hence the restriction above can be
expressed as,
\begin{equation}\label{E2cona}
    0.0817 w \leq \alpha \leq 1 \ , \ w=[10^{-2}, 12.24].
\end{equation}
The range of $w$ in both of the E-Models have been narrowed down
by three orders of magnitude, which is an implication of the
presence of the parameter $\alpha$ in our theories. A visual
representation of the above is given by the plots in Fig. \ref{E2nspl} that have been constructed for $N \simeq 60$ and are
restricted to contain the values of $n_s$ and $r$ in range with
the Planck constraints.

\subsection{Hilltop Quadratic Model}

\begin{figure}
\centering
\includegraphics[width=18pc]{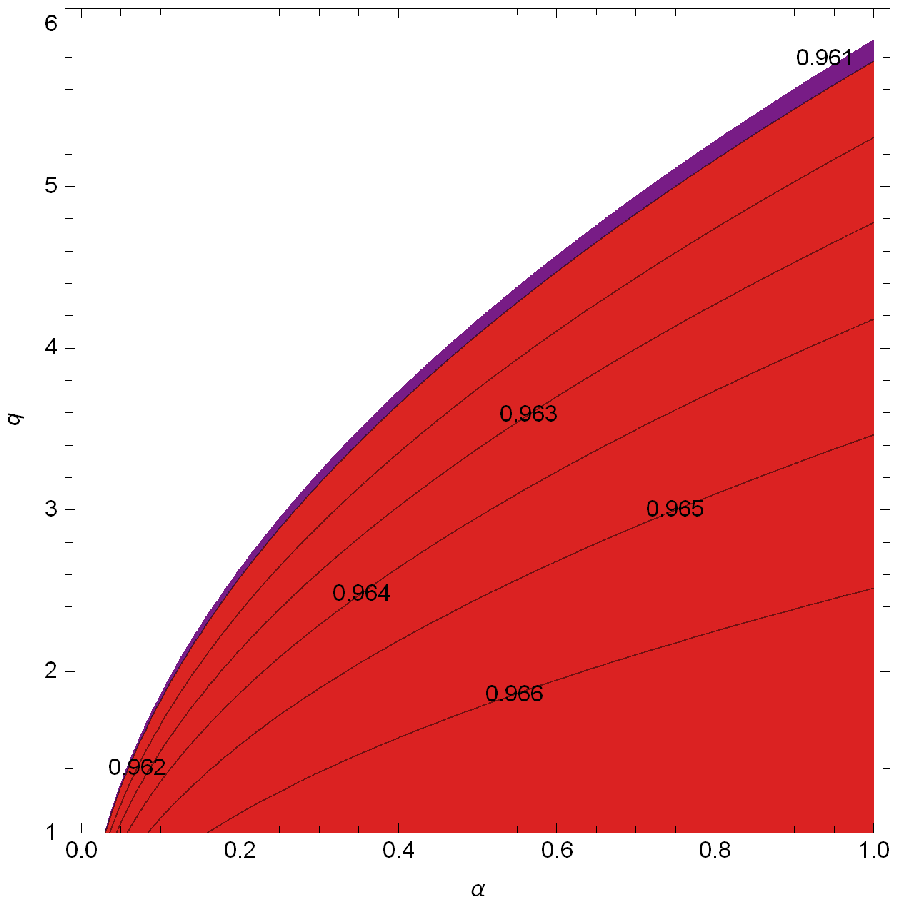}
\includegraphics[width=18pc]{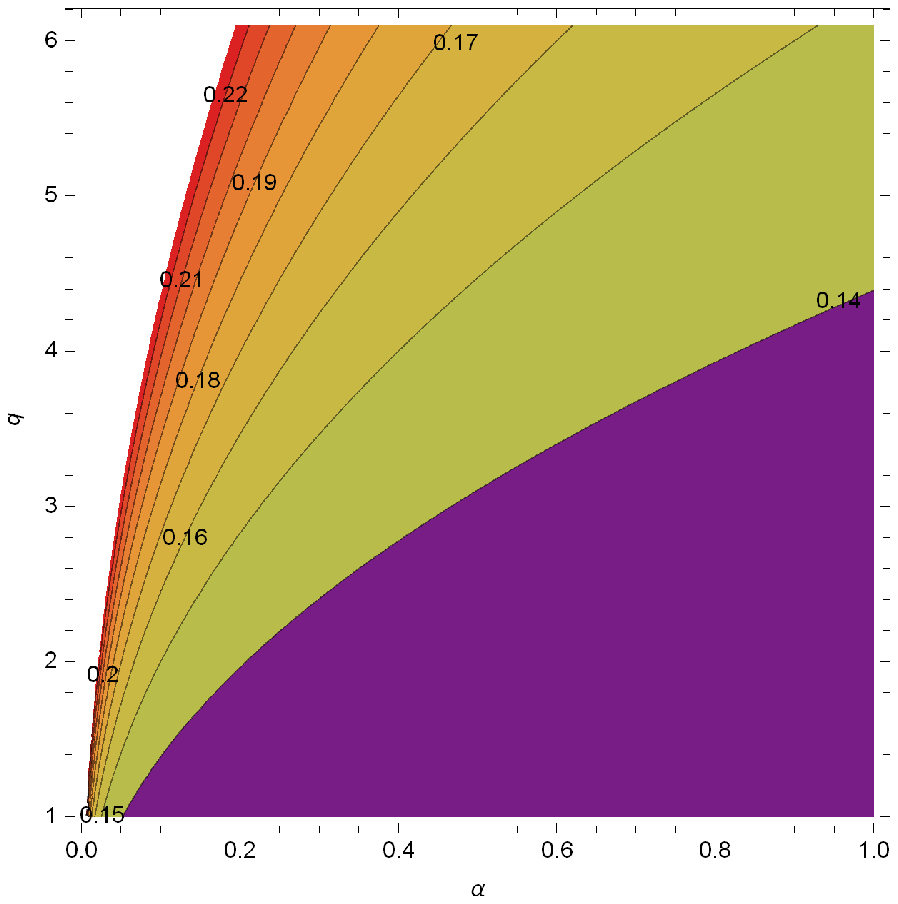}
\caption{Contour plot for the spectral index of primordial scalar
curvature perturbations $n_s$ (left plot) and the tensor-to-scalar
ratio $r$ (right plot) for $\alpha=[0, 1]$, $q=[1, 6.1]$ and
$N=60$ for the Hilltop quadratic model.}\label{Hilnspl}
\end{figure}

The Hilltop quadratic model is described by the potential,
\begin{equation}\label{HilV}
    V(\phi)=\Lambda^4 \bigg(1-\frac{\kappa^2 \phi ^2}{q ^2}\bigg),
\end{equation}
hence, the first slow-roll index $\epsilon_1$ for this potential
takes the following form,
\begin{equation}\label{Hile1}
    \epsilon_1=\frac{2 \alpha  \kappa^2 \phi ^2}{\left(q ^2-\kappa^2 \phi ^2\right)^2}.
\end{equation}
Moving on to finding $\phi_f$, the value of $\phi$ at the end of
inflation we have to solve the equation $\epsilon_1(\phi_f)=1$ and
by doing so we find that,
\begin{equation}\label{Hilff}
    \phi_f=\sqrt{\frac{\sqrt{\alpha  \kappa^4 \left(\alpha +2 q ^2\right)}}{\kappa^4}+\frac{\alpha }{\kappa^2}+\frac{q ^2}{\kappa^2}},
\end{equation}
and from the integral that gives the number of e-foldings $N$ ,
that is (\ref{N}), by solving with respect to $\phi_i$ we get,
\begin{equation}\label{Hilfi}
    \phi_i=\frac{\sqrt{\alpha +\frac{\sqrt{\alpha  \kappa^4 \left(\alpha +2 q ^2\right)}}{\kappa^2}+q ^2+4 \alpha  N}}{\kappa}.
\end{equation}
Thus, the spectral index and the tensor-to-scalar ratio to leading
order in $1/N$ are,
\begin{equation}\label{Hilns}
    n_s \simeq \frac{2 \sqrt{\alpha  \left(\alpha +2 q^2\right)}-3 q ^2+\alpha  \left(4 N^2-8 N+2\right)}{4 \alpha  N^2},
\end{equation}
\begin{equation}\label{Hilr}
   r \simeq -\frac{2 \left(\sqrt{\alpha  \left(\alpha +2 q ^2\right)}+\alpha -q ^2-4 \alpha  N\right)}{\alpha  N^2}.
\end{equation}
The viability of this model depends on the values of $\alpha$ and
$q$ for which $n_s$ and $r$ satisfy the Planck constraints in
(\ref{PlankConstraints}). For $N \simeq 60 $ the constraint on
$n_s$ is satisfied for,
 \begin{equation}\label{Hilcon}
     \frac{\alpha}{q^2} \geq 0.0286389.
 \end{equation}
We also remind that, generally, $q=[10^{0.3}, 10^{4.85}]$ and
$\alpha=[0, 1]$ and these restrictions combined with the one of
(\ref{Hilcon}) give,
 \begin{equation}\label{Hilcona}
     0.0286 \mu^2 \leq \alpha \leq 1 \ , \ q=[10^{0.3},5.9091].
 \end{equation}
However, there are no positive values for $\alpha$ and $q$ that
give $r<0.056$. These results are better presented in Fig.
\ref{Hilnspl} for $N \simeq 60$. The plot of $n_s$ has been
bounded to contain the region that $n_s=0.9649 \pm 0.0042$. As for
the plot of the tensor-to-scalar ratio $r$, it clearly shows that
$r>0.056$. Since the second Planck constraint cannot be satisfied,
these model is not viable in our inflationary theory.

\subsection{Power-law Potential $\propto$ $\phi^{2/3}$}

\begin{figure}
\centering
\includegraphics[width=18pc]{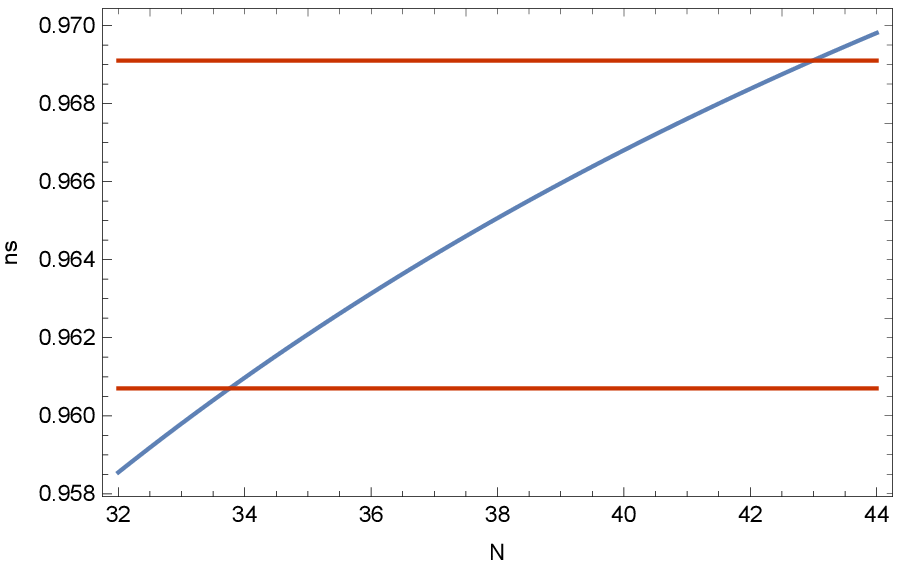}
\includegraphics[width=18pc]{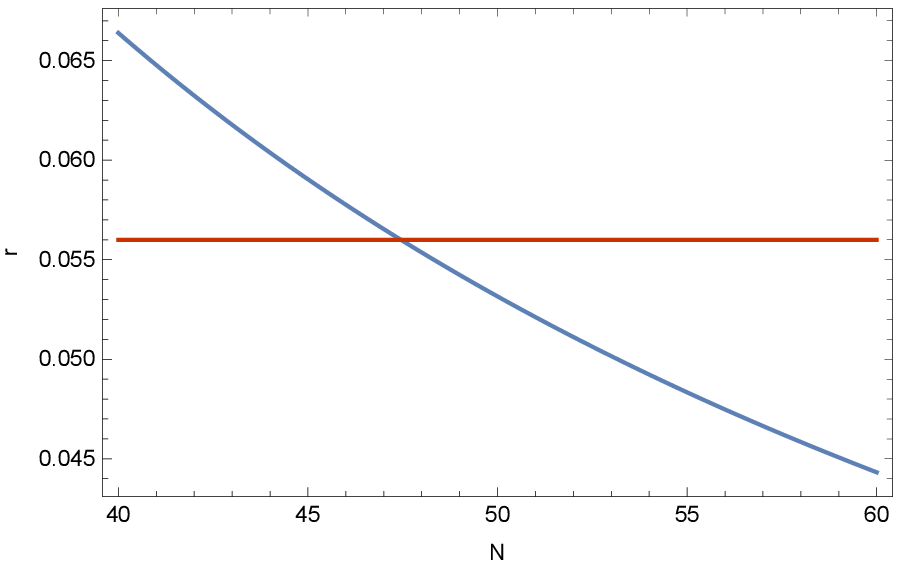}
\caption{Plots for the spectral index of primordial scalar
curvature perturbations $n_s$ for $N=[32, 44$] (left plot) and the
tensor-to-scalar ratio $r$ for $N=[40, 60]$ (right plot) for the
Power-law Potential $\propto$ $\phi^{2/3}$. The orange lines
denote the constraints $n_s=0.9649 \pm 0.0042$ and $r<0.056$
.}\label{Pl1pl}
\end{figure}

We will conclude this study of which models and under which
conditions satisfy the Planck constraints stated in
(\ref{PlankConstraints}) with the simplest models, the power-law
potentials. Their simplicity lies in the fact that their spectral
index of the primordial curvature perturbations and
tensor-to-scalar ratio are independent of the parameter $\alpha$
of our $\alpha R$ gravity model and of the potential's parameter,
leaving them dependent only on the number of $e$-foldings, $N$. We
will begin with the following potential,
\begin{equation}\label{plV1}
    V(\phi)=\frac{\lambda  \phi ^{2/3}}{\kappa^{10/3}},
\end{equation}
where $\lambda$ is a dimensionless parameter. The first slow-roll
index of this model is,
\begin{equation}\label{pl1e1}
    \epsilon_1=\frac{2 \alpha }{9 \kappa^2 \phi ^2},
\end{equation}
and by solving the equation $\epsilon_1(\phi_f)=1$ we find that,
\begin{equation}\label{pl1ff}
    \phi_f=\frac{\sqrt{2} \sqrt{\alpha }}{3 \kappa},
\end{equation}
and from (\ref{N}) we obtain the expression of $\phi_i$,
\begin{equation}\label{pl1fi}
    \phi_i=\frac{\sqrt{2} \sqrt{\alpha +6 \alpha  N}}{3 \kappa}.
\end{equation}
Now, the spectral index and the tensor-to-scalar ratio calculated
at leading order in $1/N$ are, respectively,
\begin{equation}\label{pl1ns}
    n_s \simeq \frac{2}{9 N^2}-\frac{4}{3 N}+1 ,
\end{equation}
\begin{equation}\label{pl1r}
   r \simeq \frac{8}{3 N}-\frac{4}{9 N^2}.
\end{equation}
Their independence of $\alpha$ and $\lambda$ is obvious. Hence, we
will search for the values of $N$ that generate inflationary
viability for this model based on the Planck constraints in
(\ref{PlankConstraints}) and we will discuss whether these values
of $N$ are satisfying for an inflationary model. We remind that
the inflationary era is believed to have ended at $N \simeq 50-60$
and very less probably at $N \simeq 40$. The plots of Fig. \ref{Pl1pl} show the values of $n_s$ and $r$ for different values
of $N$ with the orange lines indicating the limiting values of the
constraints. In order to obtain values of $n_s=0.9649 \pm 0.0042$
we need $N=[33.76, 42,98]$ and to have $r<0.056$ we need
$N>47.45$. Thus, there is no value of the $e$-foldings number that
can render this model viable according to these constraints.

\subsection{Power-law Potential $\propto$ $\phi$}

\begin{figure}
\centering
\includegraphics[width=18pc]{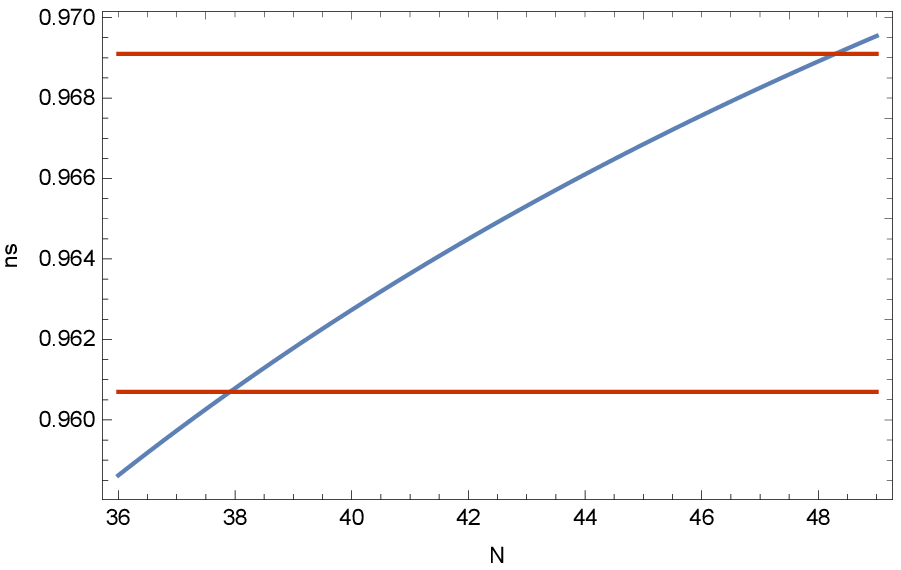}
\includegraphics[width=18pc]{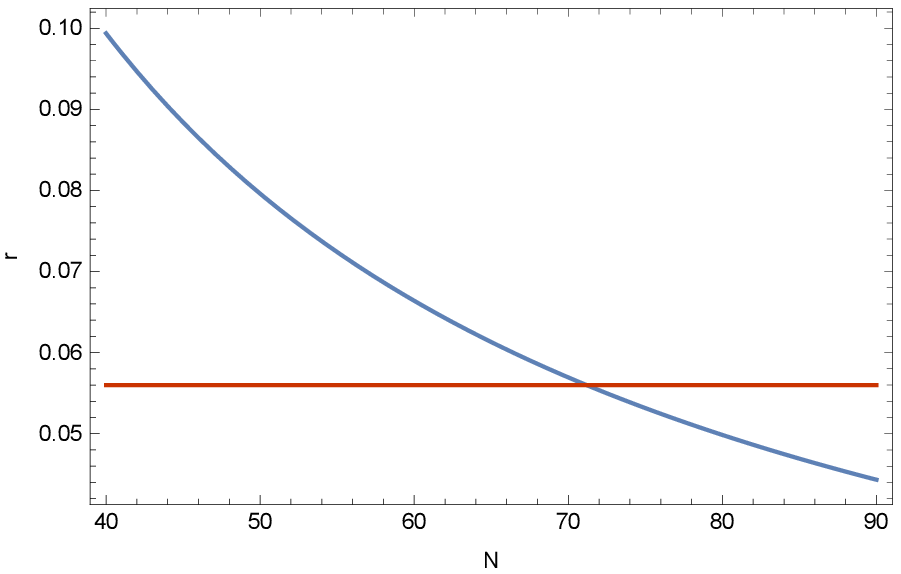}
\caption{Plots for the spectral index of primordial scalar curvature perturbations $n_s$ for $N=[36, 49]$ (left plot) and the tensor-to-scalar ratio $r$ for $N=[40, 90]$ (right plot) for the Power-law Potential $\propto$ $\phi$. The orange lines denote the constraints $n_s=0.9649 \pm 0.0042$ and $r<0.056$ .}\label{Pl2pl}
\end{figure}

The second power-law potential to study is the following:
\begin{equation}\label{pl2V}
    V(\phi)=\frac{\lambda  \phi }{\kappa^3},
\end{equation}
where $\lambda$ is again a dimensionless parameter and the first
slow-roll index for this potential is,
\begin{equation}\label{pl2e1}
    \epsilon_1=\frac{\alpha }{2 \kappa^2 \phi ^2}.
\end{equation}
From $\epsilon_1(\phi_f)=1$ we get,
\begin{equation}\label{pl2ff}
    \phi_f=-\frac{\sqrt{\alpha }}{\sqrt{2} \kappa},
\end{equation}
and solving (\ref{N}) with respect to $\phi_i$ we find,
\begin{equation}\label{pl2fi}
    \phi_i=\frac{\sqrt{\alpha } \sqrt{4 N+1}}{\sqrt{2} \kappa}.
\end{equation}
Therefore, the spectral index of the primordial curvature
perturbations and the tensor-to-scalar ratio calculated to leading
order in 1/N are,
\begin{equation}\label{pl2ns}
    n_s \simeq \frac{3}{8 N^2}-\frac{3}{2 N}+1 ,
\end{equation}
\begin{equation}\label{pl2r}
   r \simeq \frac{4 N-1}{N^2} .
\end{equation}
The plots of Fig. \ref{Pl2pl} present the values of $n_s$ and $r$
as functions of $N$ as well as the boundaries of the Planck
constraints in (\ref{PlankConstraints}). To comply with the
constraint on $n_s$, $N$ should take values in $[37.91, 48.29]$,
some of which are fairly close to the desired ones, while in order
to satisfy the constraint $r<0.056$ we need $N>71.17$, values well
outside the desired range. It is obvious that there is no value of
$N$ that can satisfy both of them simultaneously so this model is
not viable in our theory.

\subsection{Power-law Potential $\propto$ $\phi^{4/3}$}

\begin{figure}
\centering
\includegraphics[width=18pc]{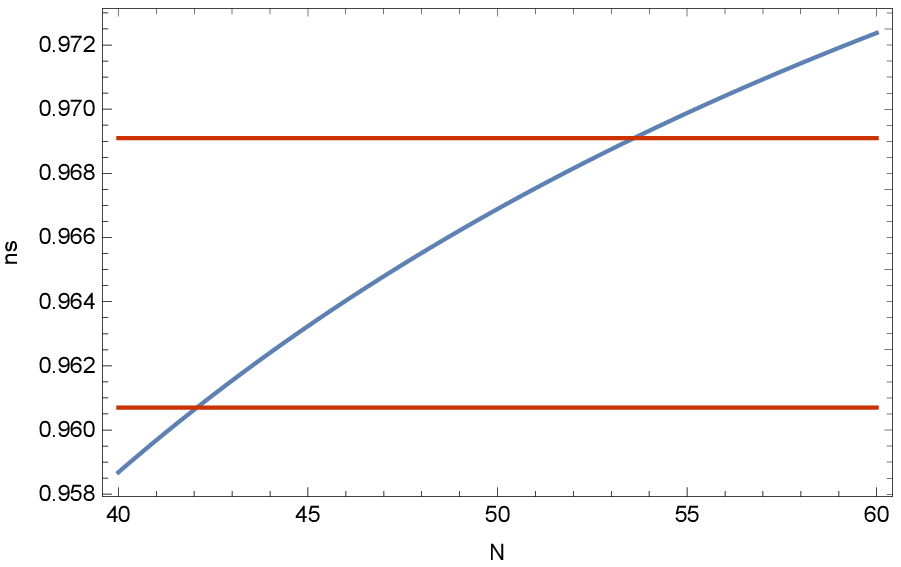}
\includegraphics[width=18pc]{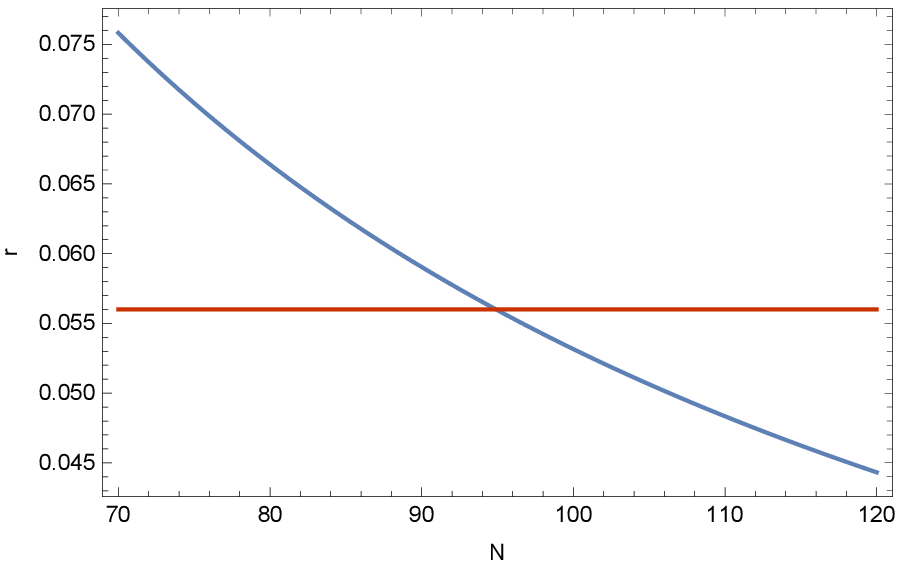}
\caption{Plots for the spectral index of primordial scalar
curvature perturbations $n_s$ for $N=[40, 60]$ (left plot) and the
tensor-to-scalar ratio $r$ for $N=[70, 120]$ (right plot) for the
power-law Potential $\propto$ $\phi^{4/3}$. The orange lines
denote the constraints $n_s=0.9649 \pm 0.0042$ and $r<0.056$
.}\label{Pl3pl}
\end{figure}

Yet another power-law model is described by the potential,
\begin{equation}\label{pl3V}
    V(\phi)=\frac{\lambda  \phi ^{4/3}}{\kappa^{8/3}},
\end{equation}
and its first slow-roll parameter is,
\begin{equation}\label{pl3e1}
    \epsilon_1=\frac{8 \alpha }{9 \kappa^2 \phi ^2}.
\end{equation}
From the equation $\epsilon_1(\phi_f)=1$ we find that,
\begin{equation}\label{pl3ff}
    \phi_f=\frac{2 \sqrt{2} \sqrt{\alpha }}{3 \kappa}
\end{equation},
and from (\ref{N}) it is easy to find $\phi_i$,
\begin{equation}\label{pl3fi}
    \phi_i=\frac{2 \sqrt{2} \sqrt{\alpha +3 \alpha  N}}{3 \kappa}.
\end{equation}
Proceeding to deriving the expressions of the spectral index $n_s$
and the tensor-to-scalar ratio $r$ we find,
\begin{equation}\label{pl3ns}
    n_s \simeq \frac{5}{9 N^2}-\frac{5}{3 N}+1 ,
\end{equation}
\begin{equation}\label{pl3r}
   r \simeq \frac{16 (3 N-1)}{9 N^2}.
\end{equation}
Searching for the values of $N$, since both $n_s$ and $r$ depend
only on this and have to satisfy the constraints $n_s=0.9649 \pm
0.0042$ and $r<0.056$ simultaneously, we find that there are no
such values. Specifically, we find out that we need $N=[42.07,
53.60]$ and $N>94.90$ respectively. These results can also be
observed in Fig.\ref{Pl3pl}. The insufficiency of the inflationary
phenomenology of this model to comply with the Planck constraints
is evident.

\subsection{Power-law Potential $\propto$ $\phi^{2}$}

The fourth power-law model to examine is described by the
potential,
\begin{equation}\label{pl4V}
    V(\phi)=\frac{\lambda  \phi ^2}{k^2}\, ,
\end{equation}
so basically the chaotic inflation model, and its first slow-roll
index expression is,
\begin{equation}\label{pl4e1}
    \epsilon_1=\frac{2 \alpha }{k^2 \phi ^2}.
\end{equation}
This slow-roll index by the end of the inflationary era becomes
equal to unity and so $\phi_f$ is,
\begin{equation}\label{pl4ff}
    \phi_f=\frac{\sqrt{2} \sqrt{\alpha }}{\kappa}.
\end{equation}
Solving (\ref{N}) with respect to $\phi_i$ we get,
\begin{equation}\label{pl4fi}
    \phi_i=\frac{\sqrt{2} \sqrt{\alpha +2 \alpha  N}}{\kappa}.
\end{equation}
Therefore, the spectral index of the primordial curvature
perturbations $n_s$ and the tensor-to-scalar ratio $r$ are
expressed as,
\begin{equation}\label{pl4ns}
    n_s \simeq \frac{(N-1)^2}{N^2},
\end{equation}
\begin{equation}\label{pl4r}
   r \simeq \frac{8 N-4}{N^2}.
\end{equation}
Beginning with the constraint $n_s=[0.9607, 0.9691]$ we find that
$N$ should belong in [50.39, 64.22] and considering that inflation
is believed to have ended at $N=[50,60]$, there are lots of values
that satisfy this constraint and belong in the desired range. That
is not the case for the tensor-to-scalar ratio that in order to
satisfy the constraint $r<0.056$ needs $N>142.36$, but such
numbers cannot stand in an inflationary theory and also there is
no accordance between the values of $N$ that satisfy each
constraint. Hence, this is another Power-law model that does not
fit in our inflationary theory.
\begin{figure}
\centering
\includegraphics[width=18pc]{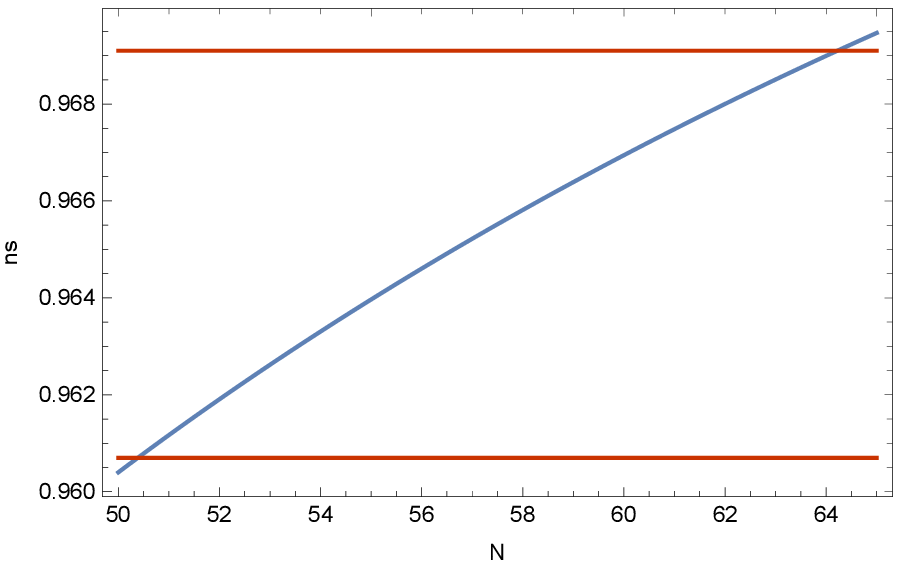}
\includegraphics[width=18pc]{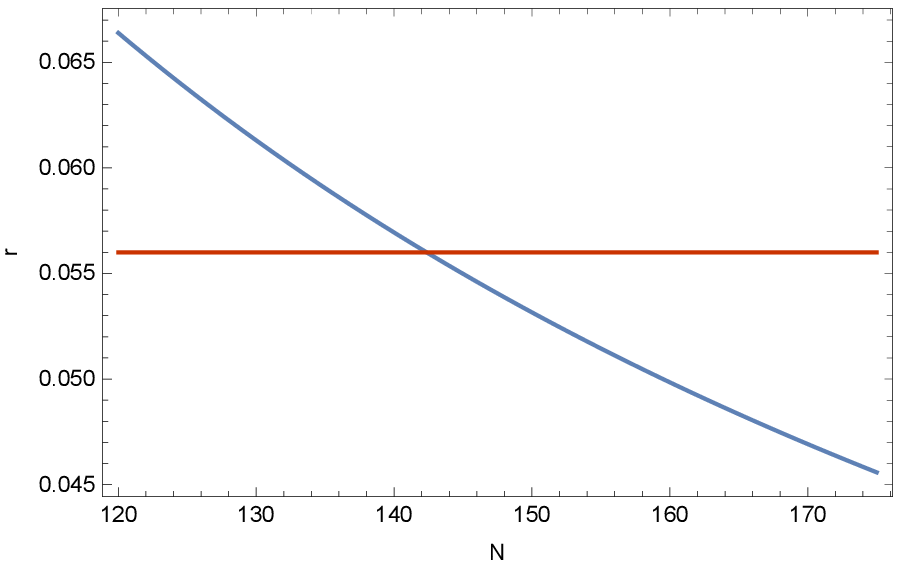}
\caption{Plots for the spectral index of primordial scalar
curvature perturbations $n_s$ for $N=[50,65]$ (left plot) and the
tensor-to-scalar ratio $r$ for $N=[120, 175]$ (right plot) for the
Power-law Potential $\propto$ $\phi^{2}$. The orange lines denote
the constraints $n_s=0.9649 \pm 0.0042$ and $r<0.056$
.}\label{Pl4pl}
\end{figure}

\subsection{Power-law Potential $\propto$ $\phi^{3}$}

\begin{figure}
\centering
\includegraphics[width=18pc]{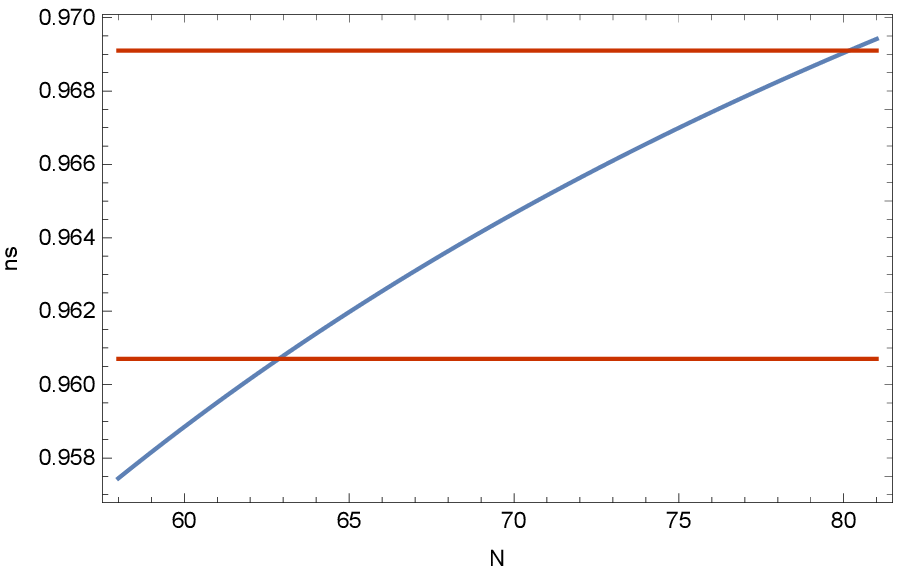}
\includegraphics[width=18pc]{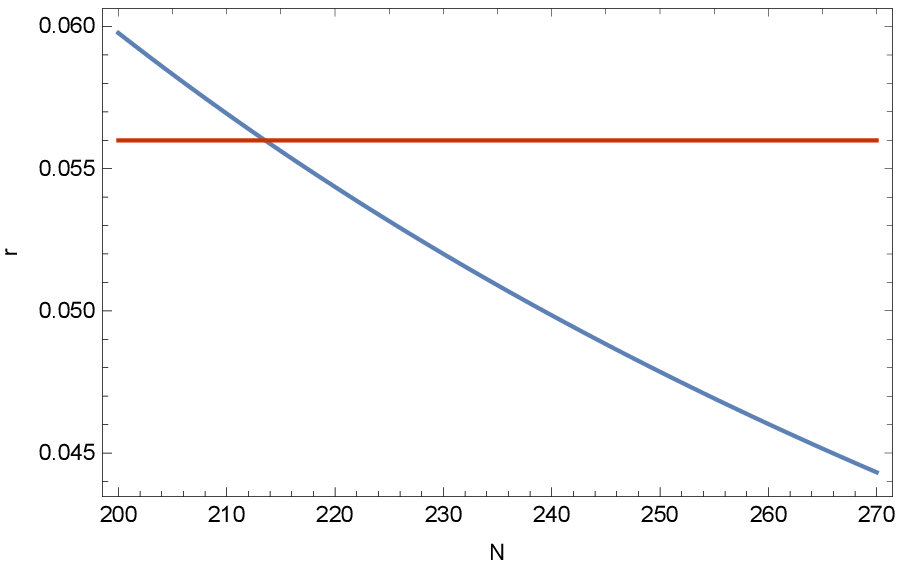}
\caption{Plots for the spectral index of primordial scalar
curvature perturbations $n_s$ for $N=[58, 81]$ (left plot) and the
tensor-to-scalar ratio $r$ for $N=[200, 270]$ (right plot) for the
power-law Potential $\propto$ $\phi^{3}$. The orange lines denote
the constraints $n_s=0.9649 \pm 0.0042$ and $r<0.056$
.}\label{Pl5pl}
\end{figure}

We proceed our study with another power-law potential,
\begin{equation}\label{pl5V}
    V(\phi)=\frac{\lambda  \phi ^3}{\kappa}.
\end{equation}
The first slow-roll index of this potential is,
\begin{equation}\label{pl5e1}
    \epsilon_1=\frac{9 \alpha }{2 k^2 \phi ^2}
\end{equation}
and at the end of inflation it holds true that $\epsilon_1(\phi_f)=1$, thus,
\begin{equation}\label{pl5ff}
    \phi_f=\frac{3 \sqrt{\alpha }}{\sqrt{2} \kappa},
\end{equation}
and
\begin{equation}\label{pl5fi}
    \phi_i=\frac{\sqrt{\frac{3}{2}} \sqrt{3 \alpha +4 \alpha  N}}{\kappa}.
\end{equation}
When it comes to the spectral index and the tensor-to-scalar
ratio, they are calculated at leading order in $1/N$ and found
equal to,
\begin{equation}\label{pl5ns}
    n_s\simeq\frac{4 N-7}{4 N+3},
\end{equation}
\begin{equation}\label{pl5r}
   r\simeq\frac{3 (4 N-3)}{N^2}.
\end{equation}
Both $n_s$ and $r$ depend only on the number of $e$-foldings, $N$,
and in order to satisfy the Planck constraints in
(\ref{PlankConstraints}) we need $N=[62.86, 80.16]$ and $N>213.53$
respectively. It is obvious that there is no value of $N$ such
that both indices are within the constraints range at once.
Furthermore, these values exceed the initial range of $N$ that is
$[50, 60]$. A better visualization of these results is provided by
the plots of Fig. \ref{Pl5pl}.

\subsection{Power-law Potential $\propto$ $\phi^{4}$}

The last Power-law potential we will examine is the following,
\begin{equation}\label{pl6V}
    V(\phi)=\lambda  \phi ^4 .
\end{equation}
The first slow-roll index of this potential is,
\begin{equation}\label{pl6e1}
    \epsilon_1=\frac{8 \alpha }{\kappa^2 \phi ^2},
\end{equation}
and from $\epsilon_1(\phi_f)=1$ we obtain $\phi_f$,
\begin{equation}\label{pl6ff}
    \phi_f=\frac{2 \sqrt{2} \sqrt{\alpha }}{\kappa}.
\end{equation}
The value of $\phi$ at the beginning of inflation can be derived
from (\ref{N}) and is found to be,
\begin{equation}\label{pl6fi}
    \phi_i=\frac{2 \sqrt{2} \sqrt{\alpha +\alpha  N}}{\kappa}.
\end{equation}
Moving on as usual, the spectral index and the tensor-to-scalar
ratio are,
\begin{equation}\label{pl6ns}
    n_s \simeq \frac{3}{N^2}-\frac{3}{N}+1 ,
\end{equation}
\begin{equation}\label{pl6r}
   r \simeq 16 \left(\frac{1}{N}-\frac{1}{N^2}\right) .
\end{equation}
Demanding $n_s$ and $r$ to satisfy the Planck constraints in
(\ref{PlankConstraints}), we reach some restrictions on the values
that $N$ can take which do not belong in the range $N=[50, 60]$.
Specifically, $n_s$ complies with its constraint when $N=[75.32,
96.08]$ and $r$ when $N>284.71$ so, the constraints cannot be
satisfied simultaneously for this model as well.
\begin{figure}
\centering
\includegraphics[width=18pc]{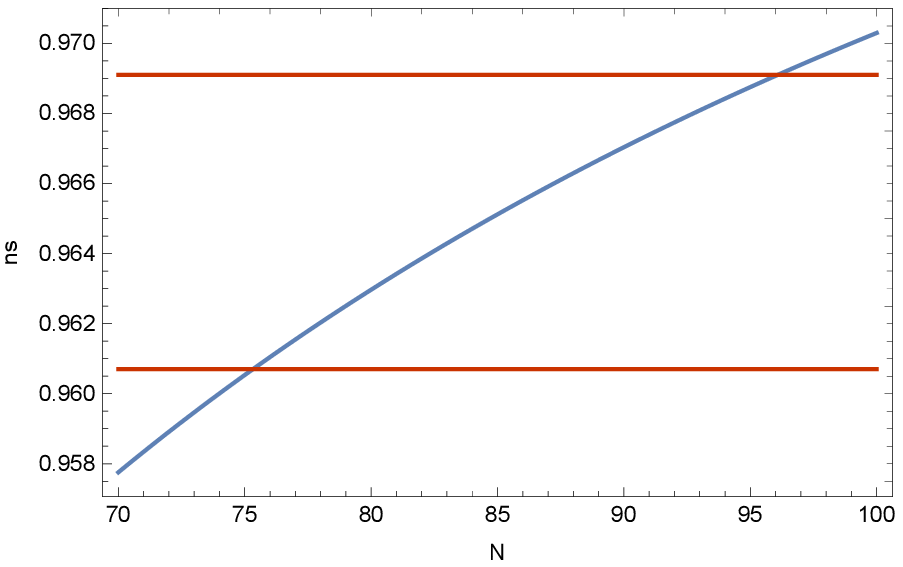}
\includegraphics[width=18pc]{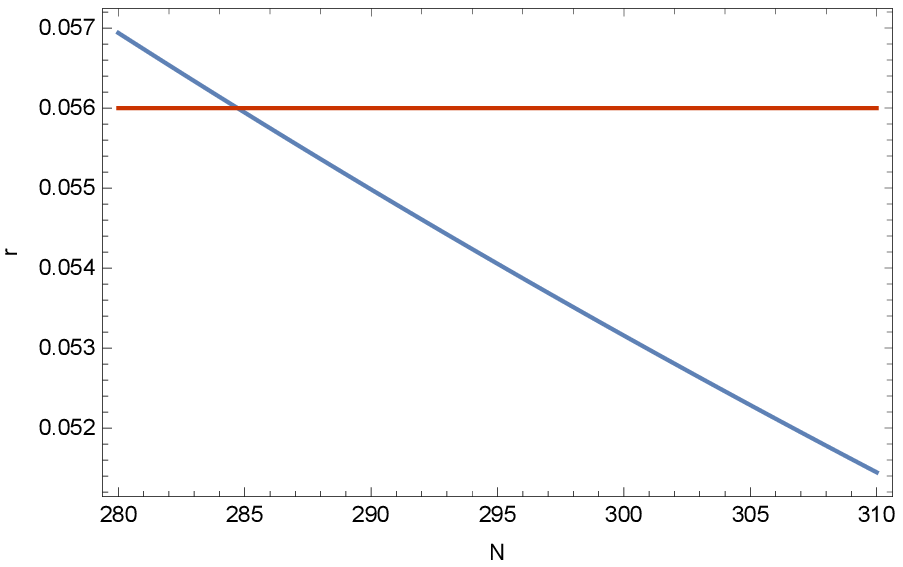}
\caption{Plots for the spectral index of primordial scalar
curvature perturbations $n_s$ for $N=[70, 100]$ (left plot) and
the tensor-to-scalar ratio $r$ for $N=[280, 310]$ (right plot) for
the Power-law Potential $\propto$ $\phi^{4}$. The orange lines
denote the constraints $n_s=0.9649 \pm 0.0042$ and $r<0.056$
.}\label{Pl6pl}
\end{figure}
In the next section, we will examine whether some of the models we
studied also satisfy the Swampland criteria, under which
conditions and at the same time as the Planck constraints.


\section{Swampland Criteria for Several Rescaled Einstein-Hilbert Gravity Models}

\subsection{Swampland Criteria for Power-law Potential $\propto$
$\phi^{2}$}

\begin{figure}
\centering
\includegraphics[width=18pc]{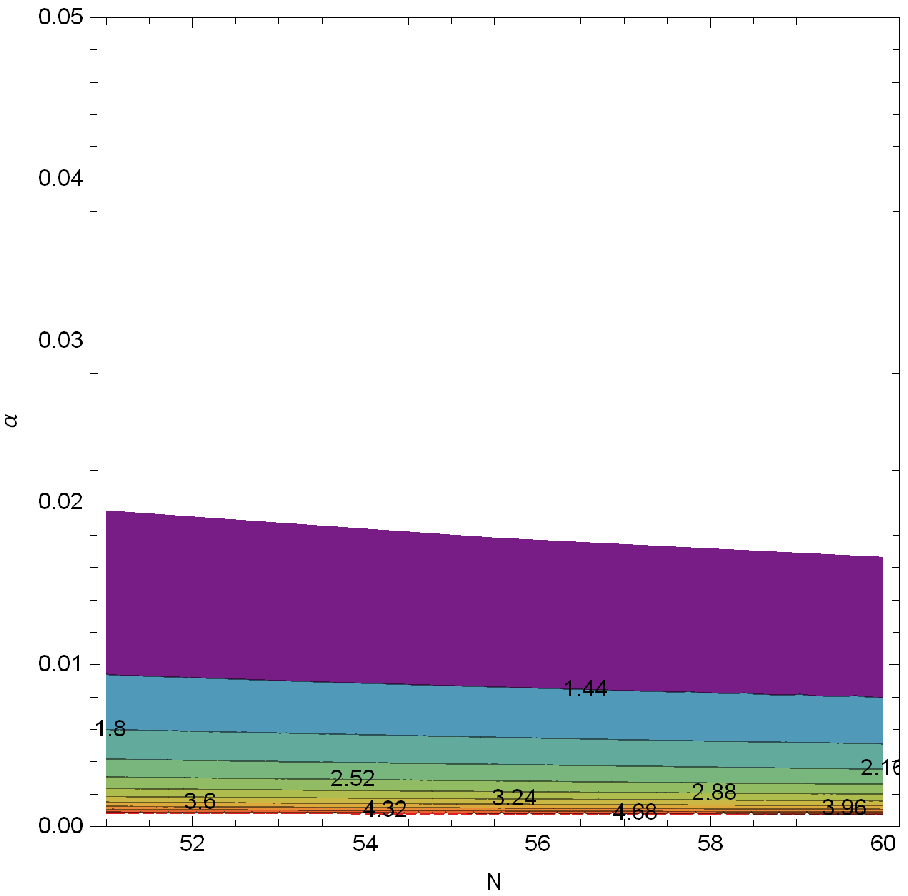}
\includegraphics[width=18pc]{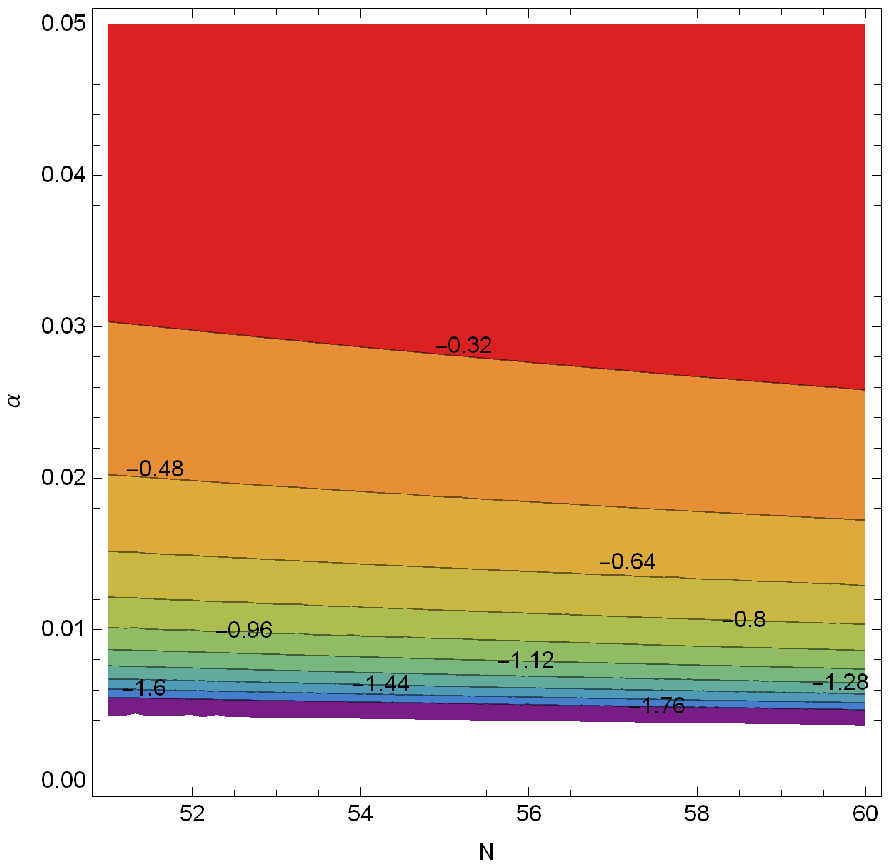}
\caption{Second and third Swampland criteria analysis for the
Power-law Potential $\propto$ $\phi^{2}$. Contour plots of
$V'(\phi)/V(\phi)$(left plot) and of $-V''(\phi)/V(\phi)$ (right
plot), evaluated at the first horizon crossing for the ranges of
values of the free parameters, N = [51,60] and $\alpha = [0,
0.05]$ in reduced Planck units.}\label{pl2sw}
\end{figure}

We proceed to calculating the values of $\alpha$ respecting the
constraints obtained by inflation. By taking into consideration
(\ref{S.C. v'/v}) and after calculations we obtain that,
\begin{equation}
    \centering\label{power v'/v with N}
    \frac{V'(\phi_i)}{V(\phi_i)}=\frac{\kappa}{\sqrt{\alpha(1/2+N)}}.
\end{equation}
For $\kappa=1$, we need to find the contour plot of (\ref{power
v'/v with N}), so we conclude to Fig. \ref{pl2sw}.
Thanks to Fig. \ref{pl2sw} we obtain that for $\alpha<0.02$
(\ref{S.C. v'/v}) is satisfied. In addition, we should check the
second Swampland criterion, and then compare both results. Bearing
in mind (\ref{S.C. v''/v}), we obtain,
\begin{equation}
    \centering\label{power v''/v with N}
    -\frac{V''(\phi_i)}{V(\phi_i)}=\frac{\kappa^2}{\alpha(1+2N)}.
\end{equation}
For $\kappa=1$, the contour plot of (\ref{power v''/v with N}) is given in Fig. \ref{pl2sw}.

So by taking into account Fig. \ref{pl2sw} and if we need to have the constraints on inflation and the Swampland criteria satisfied we need:
\begin{equation}
    \centering
    \boxed{\alpha=[0,0.01]}.
\end{equation}
The primordial tilt and the tensor-to-scalar ratio appear to be
invariant from the free parameters inserted in the potential
equation. In order to present an unabridged analysis we have
chosen to consider the $e$-foldings number as a variable.

\subsection{Swampland Criteria for a T-Model(m=1) Model}

\begin{figure}
\centering
\includegraphics[width=18pc]{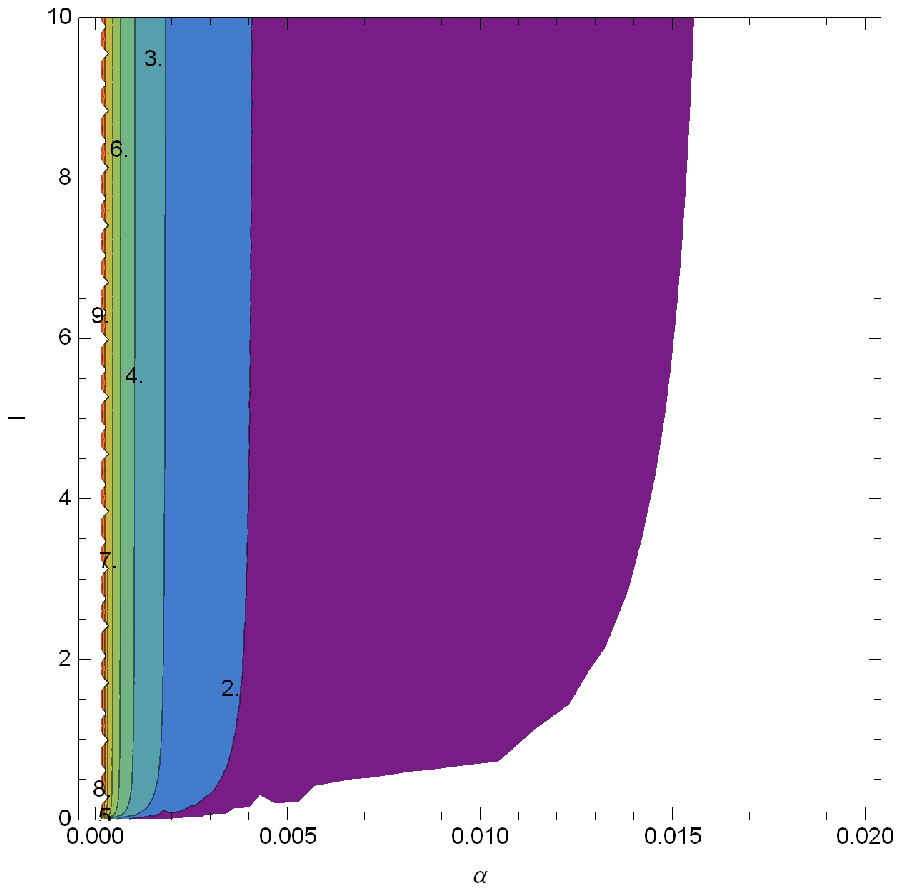}
\includegraphics[width=18pc]{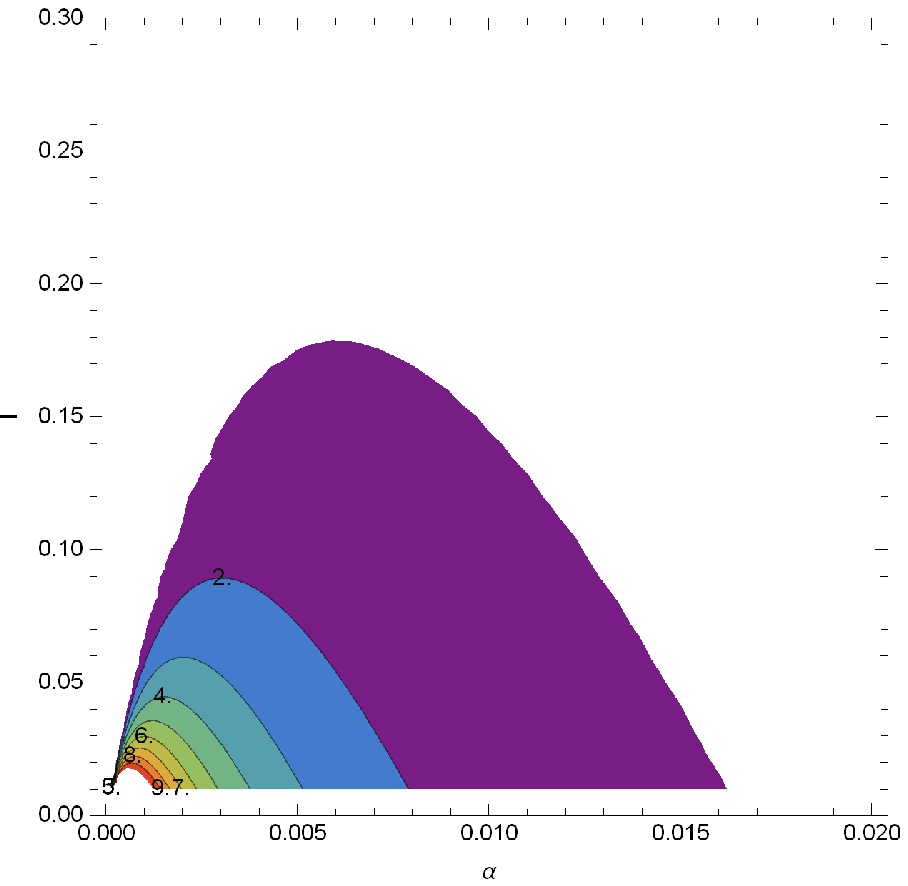}
\caption{Second and third Swampland criteria analysis for the
T-Model(m=1). Contour plots of $V'(\phi)/V(\phi)$(left plot) and
of $-V''(\phi)/V(\phi)$(right plot), evaluated at the first
horizon crossing. In both plots we set $N=60$ and we work in
reduced Planck units. For the left plot $\alpha=[0,0.02]$ and
$w=[0,10]$ and for the right plot $\alpha=[0,0.2]$ and
$w=[0.01,0.3]$.}\label{t1sw}
\end{figure}

We proceed our analysis for the T-Model (m=1) Model. At this point
we will check the values of $\alpha$ and $\lambda$ so that
(\ref{S.C. v'/v}) is satisfied  for this potential. Taking into
consideration (\ref{S.C. v'/v}) and after calculations we obtain:
\begin{equation}
    \centering\label{t model v'/v with N}
    \frac{V'(\phi_i)}{V(\phi_i)}=\frac{2 \sqrt{\frac{2}{3}} \kappa \text{csch}\left(\kappa \cosh ^{-1}\left(\frac{l \sqrt{\frac{12 \alpha}{l}+9}+4 \alpha N}{3 l}\right)\right)}{\sqrt{l}}.
\end{equation}
For $\kappa=1$ and $N=60$, we reach for (\ref{t model v'/v with
N}) Fig. \ref{t1sw}.

At this point we have to focus on the second Swampland criterion
and then extract the constraints on $\alpha$ and $l$. We also need
to bear in mind (\ref{S.C. v''/v}). But, as we are going to find
out the expression is rather complicated. So the result, by taking
into account that $\kappa=1$ and $N=60$, is,
\begin{equation}
    \centering\label{tmodel v''/v with N}
    -\frac{V''(\phi_i)}{V(\phi_i)}=-\frac{l \sqrt{\frac{12\alpha}{l}+9}+240 \alpha-6l}{3 \alpha \left(40 l \sqrt{\frac{12 \alpha}{l}+9}+4800 \alpha+l\right)}.
\end{equation}
The contour plot of (\ref{tmodel v''/v with N}) is given in Fig. \ref{t1sw}. Consequently, if we need both the constraints on
inflation and the Swampland criteria satisfied the parameter
$\alpha$ must be in the range $\alpha=[0,0.008]$

\subsection{Swampland Criteria for a D-brane inflation (p=4) Model}

\begin{figure}
\centering
\includegraphics[width=18pc]{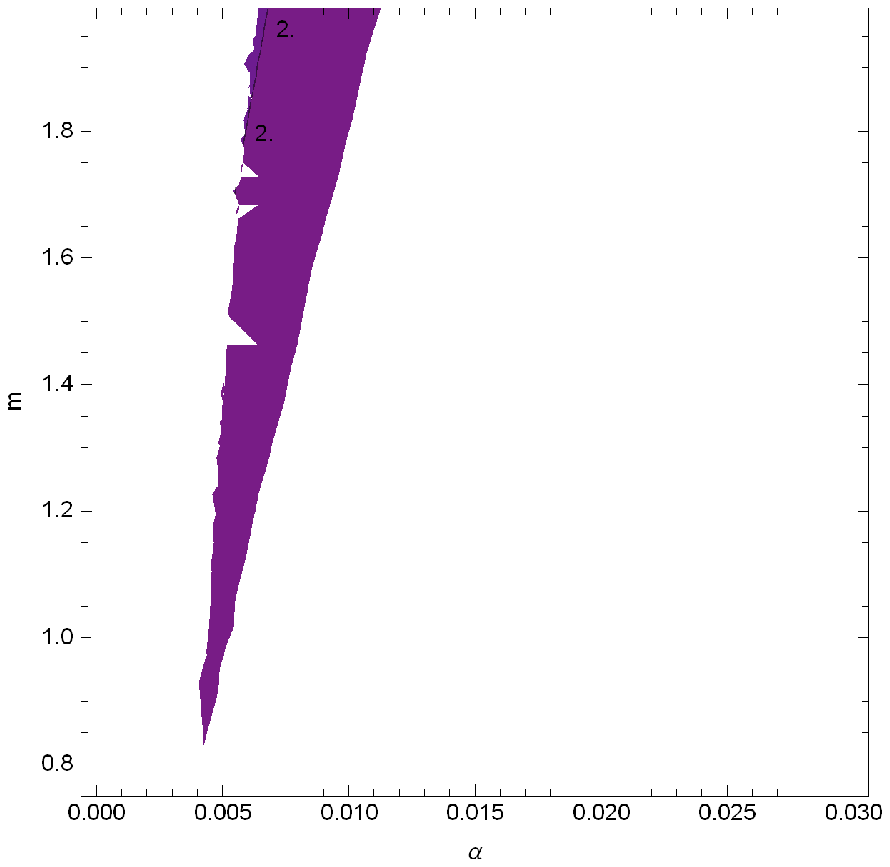}
\includegraphics[width=18pc]{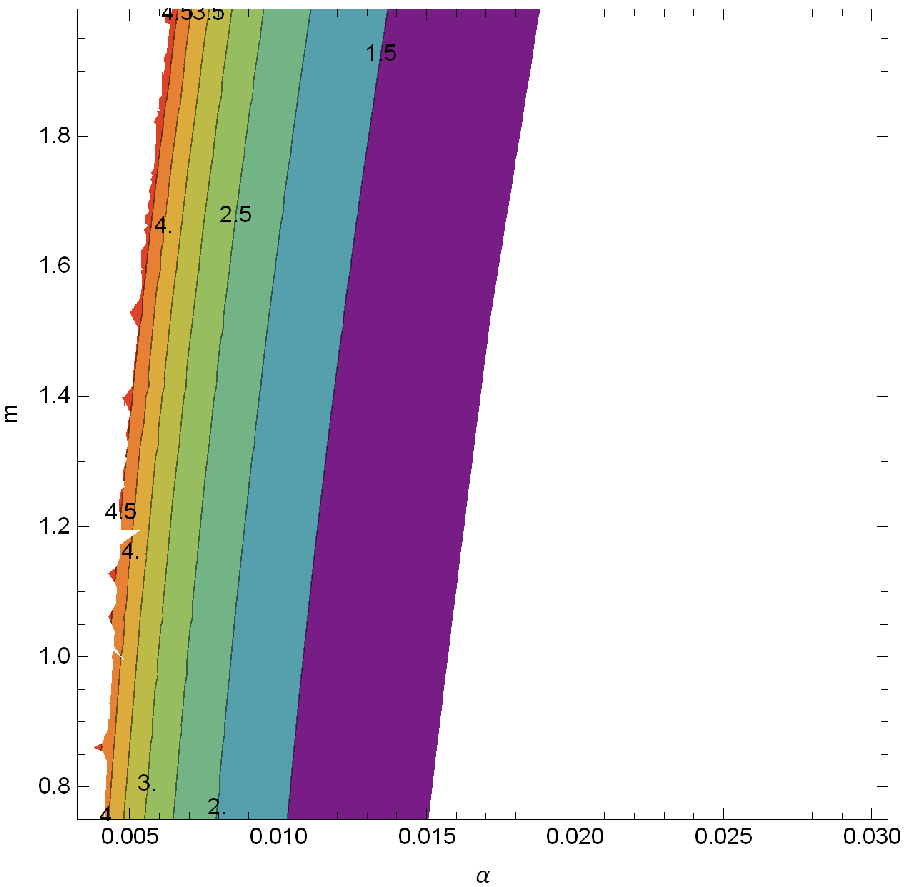}
\caption{Second and third Swampland criteria analysis for D-brane
inflation (p=4) Model. Contour plots of $V'(\phi)/V(\phi)$(left
plot) and of $-V''(\phi)/V(\phi)$(right plot), evaluated at the
first horizon crossing for the ranges of values of the free
parameters, $ m = [0.75, 1.99526]$ and $\alpha= [0, 0.03]$ in
reduced Planck units for $N = 60$.}\label{dbr4sw}
\end{figure}

At this point we check the Swampland criteria for the D-Brane
(p=4) potential, with respect to the values of $\alpha$ and
$\lambda$ that the constraints impose. Bearing in mind (\ref{S.C.
v'/v}) and after calculations we obtain:
\begin{equation}
    \centering\label{dmodel v'/v with N}
    \frac{V'(\phi_i)}{V(\phi_i)}=-\frac{4 \kappa m^4}{m^4 \sqrt[6]{2\ 2^{4/5} \alpha^{3/5} m^{24/5}+24 \alpha m^4 N}-\left(2\ 2^{4/5} \alpha^{3/5} m^{24/5}+24 \alpha m^4 N\right)^{5/6}}.
    \end{equation}
For $\kappa=1$ and $N=60$, the contour plot of (\ref{dmodel v'/v with N}) is given in Fig. \ref{dbr4sw}.
Before deducing any results, we also have to check the second Swampland criterion so, by taking into account (\ref{S.C. v''/v}) and after calculations we find:
\begin{equation}
    \centering\label{dmodel v''/v with N}
    -\frac{V''(\phi_i)}{V(\phi_i)}=\frac{20 \kappa^2}{-\sqrt[3]{2\ 2^{4/5} \alpha^{3/5} m^{24/5}+24 \alpha m^4 N}+2\ 2^{4/5} \alpha^{3/5} m^{4/5}+24 \alpha N}.
    \end{equation}
For $\kappa=1$ and $N=60$, the contour plot of (\ref{dmodel v''/v with N}) is given in Fig. \ref{dbr4sw}.

By comparing the right and left plot in Fig. \ref{dbr4sw}, we
obtain that for the following values of $\alpha$ and $m$ we have
both Swampland criteria and inflation satisfied,
\begin{equation}
\centering\label{dbrane sc results}
\boxed{
\begin{array}{rcl}
\alpha&=&[0, 0.03]\\
m&=&[0.75, 1.99526] \\
512.802\alpha-0.318&<& m\\
m&<&164.067\alpha+0.146
\end{array}}
\end{equation}

\subsection{Swampland Criteria for an E-model (n=1)}

\begin{figure}
\centering
\includegraphics[width=18pc]{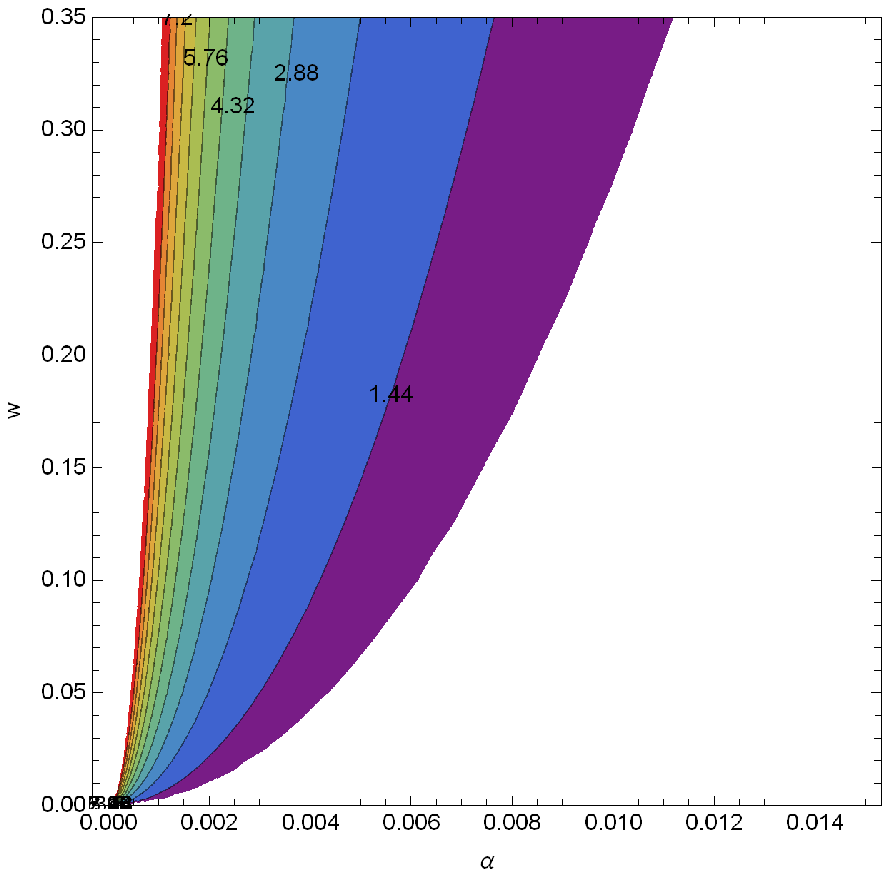}
\includegraphics[width=18pc]{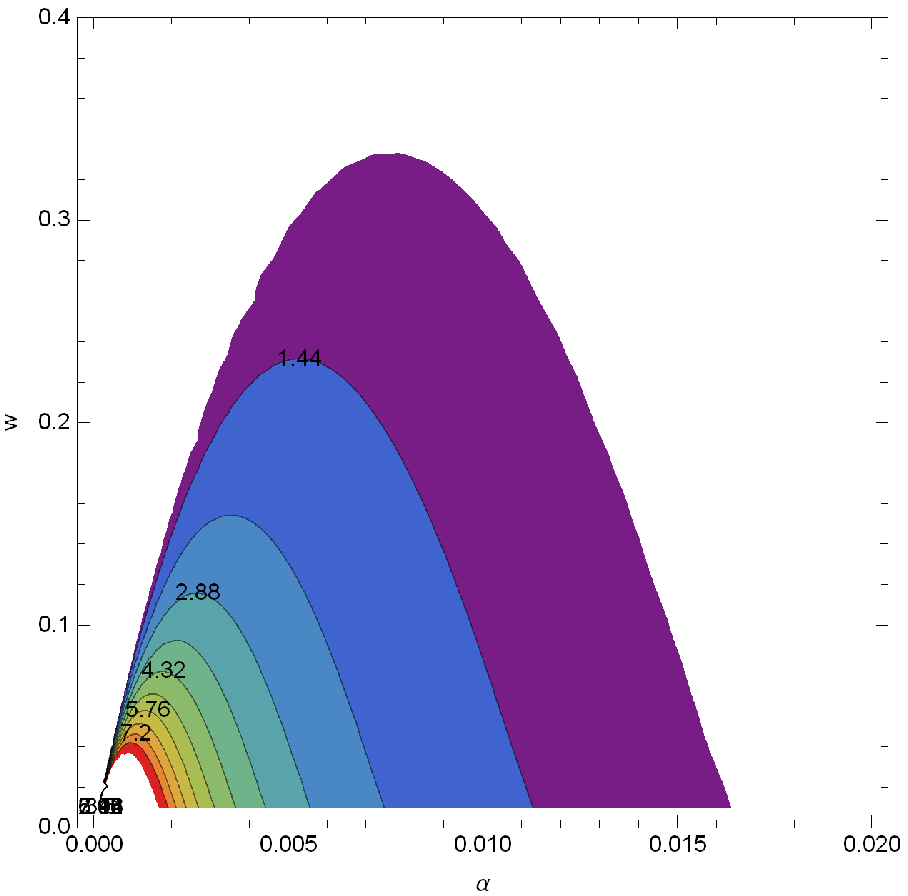}
\caption{Second and third Swampland criteria analysis for the
E-model (n=1). Contour plots of $V'(\phi)/V(\phi)$(left plot) and
of $-V''(\phi)/V(\phi)$(right plot), evaluated at the first
horizon crossing. In both plots we set $N=60$ and we work in
reduced Planck units. For the left plot $\alpha=[0,0.015]$ and
$w=[0,0.35]$ and for the right plot $\alpha=[0,0.2]$ and
$w=[0.01,0.04]$.}\label{e1sw}
\end{figure}

We proceed our work by checking the Swampland criteria for an
E-model (N=1), with respect to the values of $\alpha$ and $w$ that
the inflationary constraints impose. Bearing in mind (\ref{S.C.
v'/v}) and after calculations we obtain:
\begin{equation}
    \centering\label{emodel v'/v with N}
    \frac{V'(\phi_i)}{V(\phi_i)}=\frac{\sqrt{6} \kappa \sqrt{w}}{\sqrt{3} \sqrt{\alpha} \sqrt{w}+2 \alpha N}.
    \end{equation}
For $\kappa=1$ and $N=60$, the contour plot of (\ref{emodel v'/v
with N}) is given in Fig. \ref{e1sw}.
We also need to check the second Swampland criterion, and then reach a result so, by taking into consideration (\ref{S.C. v''/v}) and after calculations we obtain:
\begin{equation}
    \centering\label{emodel v''/v with N}
    -\frac{V''(\phi_i)}{V(\phi_i)}=\frac{\kappa^2 \left(2 \sqrt{3} \sqrt{\alpha} \sqrt{w}+4 \alpha N-3 w\right)}{\alpha \left(2 \sqrt{\alpha} N+\sqrt{3} \sqrt{w}\right)^2}.
    \end{equation}
For $\kappa=1$ and $N=60$, the contour plot of (\ref{emodel v''/v
with N}) is given in Fig. \ref{e1sw}.

Obviously, the set of values that satisfy the second Swampland
criterion is wider than the one satisfying the third one. If we
demand to have either the first or the second Swampland criterion
satisfied, we have to obtain the following sets of values for the
free parameters:
\begin{equation}
\centering\label{emodel results}
\boxed{
\begin{array}{rcl}
\alpha&=&[0, 0.05138]\\
w&=&[0.01, 0.05138] \\
w&>&3366.41\alpha^2+0.01
\end{array}}
\end{equation}


\subsection{Swampland Criteria for the Natural Inflation Model}

\begin{figure}
\centering
\includegraphics[width=18pc]{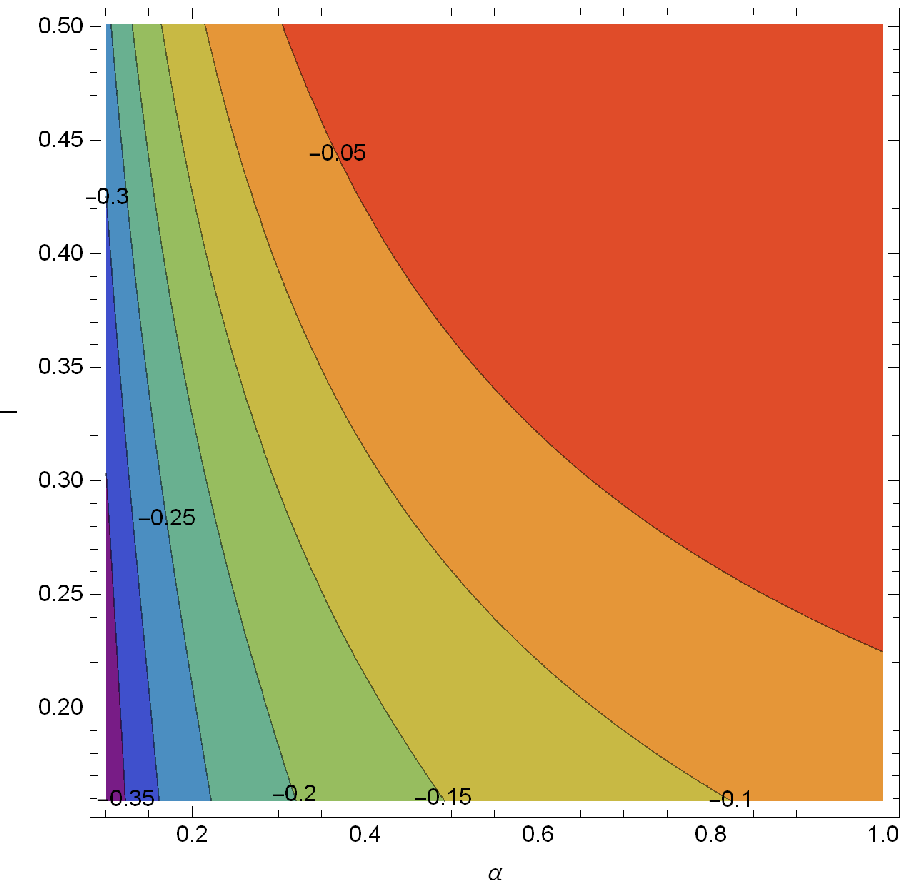}
\includegraphics[width=18pc]{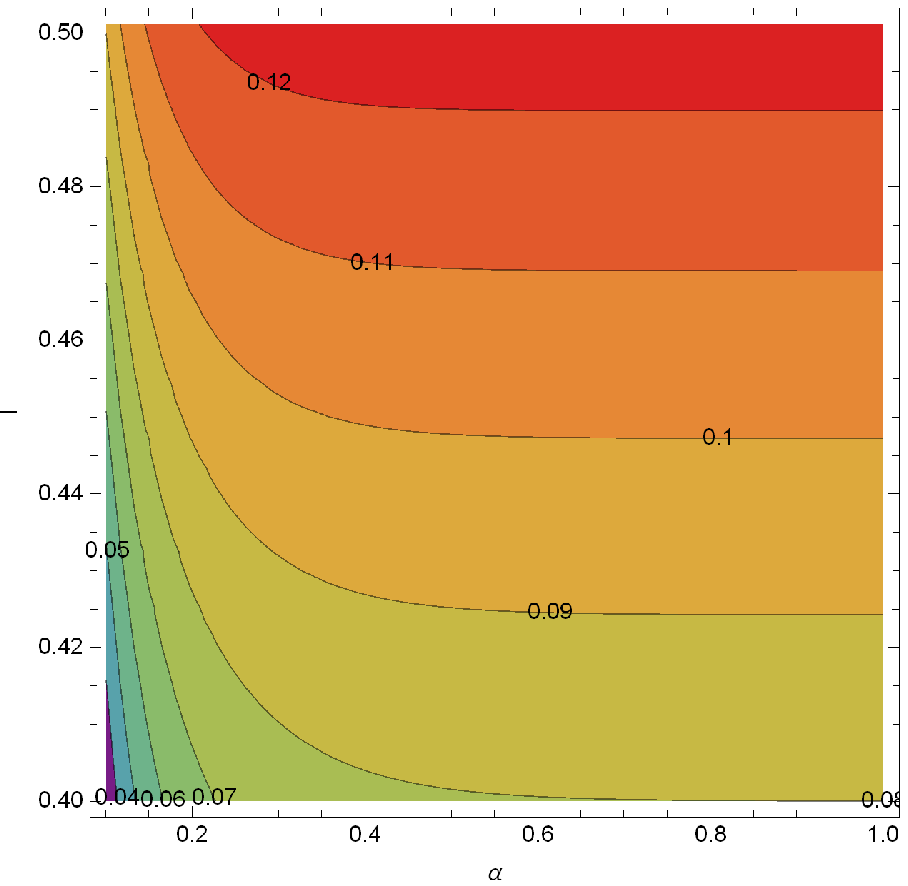}
\caption{Second and third Swampland criteria analysis for the
natural model. Contour plots of $V'(\phi)/V(\phi)$(left plot) and
of $-V''(\phi)/V(\phi)$(right plot), evaluated at the first
horizon crossing for the ranges of values of the free parameters $
l = [0.0032,0.50112]$ and $\alpha= [0.1006,1]$ in reduced Planck
units for $N = 60$.}\label{natsw12}
\end{figure}

In this subsection we check the Swampland criteria for the Natural
inflation model, with respect to the values of $\alpha$ and $w$
that the constraints impose. Taking into account (\ref{S.C. v'/v})
and after calculations we find:
\begin{equation}
    \centering\label{natural v'/v with N}
    \frac{V'(\phi_i)}{V(\phi_i)}=-\frac{l \sin \left(2 \sin ^{-1}\left(\frac{\sqrt{2} e^{-30 \alpha l^2}}{\sqrt{\alpha} l \sqrt{\frac{2}{\alpha l^2}+1}}\right)\right)}{\cos \left(2 \sin ^{-1}\left(\frac{\sqrt{2} e^{-30 \alpha l^2}}{\sqrt{\alpha} l \sqrt{\frac{2}{\alpha l^2}+1}}\right)\right)+1}.
    \end{equation}
For $\kappa=1$ and $N=60$, the plot of (\ref{natural v'/v with N})
is given in Fig. \ref{natsw12}.
Fig. \ref{natsw12} indicates that the first Swampland criterion
can not be satisfied. So, we have to proceed to the second
Swampland criterion. Bearing in mind (\ref{S.C. v''/v}), after
calculations we obtain:
\begin{equation}
    \centering\label{natural v''/v with N}
    -\frac{V''(\phi_i)}{V(\phi_i)}=\frac{l^2 \cos \left(2 \sin ^{-1}\left(\frac{\sqrt{2} e^{-30 \alpha l^2}}{\sqrt{\alpha} l \sqrt{\frac{2}{\alpha l^2}+1}}\right)\right)}{\cos \left(2 \sin ^{-1}\left(\frac{\sqrt{2} e^{-30 \alpha l^2}}{\sqrt{\alpha} l \sqrt{\frac{2}{\alpha l^2}+1}}\right)\right)+1}.
    \end{equation}
For $\kappa=1$ and $N=60$, the contour plot of (\ref{natural v''/v with N}) is given in Fig. \ref{natsw12}.

\begin{figure}
\centering
\includegraphics[width=18pc]{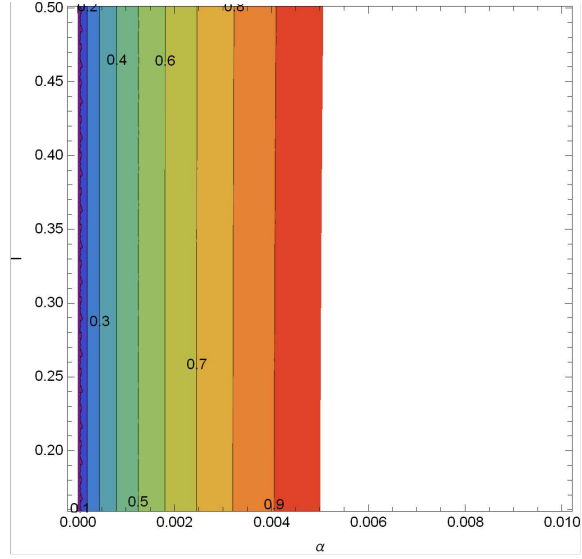}
\caption{First Swampland criterion analysis for the natural model. Contour plot of $\Delta\phi$, evaluated at the first horizon crossing for the ranges of values of the free parameters, $ l = [0.0032,0.50112]$ and $\alpha= [0.1006,1]$ in reduced Planck units for $N = 60$.}\label{natsw3}
\end{figure}

It may seem that we have reached a dead end, because none of the
Swampland criteria is satisfied. At this point, we need to recall
that $\Delta\phi\leq f\sim\mathcal{O}(1)$ so, bearing in mind the
initial and the final values of $\phi$ for the Natural inflation
potential, we obtain:
\begin{equation}
\centering\label{natural sc3 equation}
\Delta\phi=\frac{2 \tan ^{-1}\left(\frac{\sqrt{2}}{\sqrt{\alpha} l}\right)}{\kappa l}-\frac{2 \sin ^{-1}\left(\frac{\sqrt{2} e^{-\frac{1}{2} a l^2 N}}{\sqrt{\alpha} l \sqrt{\frac{2}{\alpha l^2}+1}}\right)}{\kappa l}.
\end{equation}
For $\kappa=1$ and $N=60$, the contour plot of (\ref{natural sc3 equation}) is given in Fig. \ref{natsw3}.
So, we finally find that there is no possibility that the
inflationary constraints and the Swampland criteria are satisfied
at the same time for the Natural inflation model.

\subsection{Swampland Criteria for the E - Model (N=2)}

\begin{figure}
\centering
\includegraphics[width=18pc]{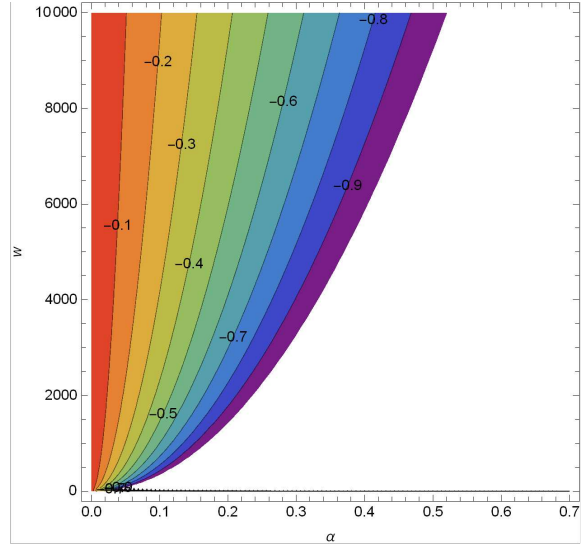}
\caption{First Swampland criterion analysis for E-model (N=2).
Contour plot of $\Delta\phi$, evaluated at the first horizon
crossing for the ranges of values of the free parameters $\alpha =
[0 , 0.7]$, $w = [0.01, 10000]$  in reduced Planck units for $N =
60$.} \label{e2sw1}
\end{figure}

\begin{figure}
\centering
\includegraphics[width=18pc]{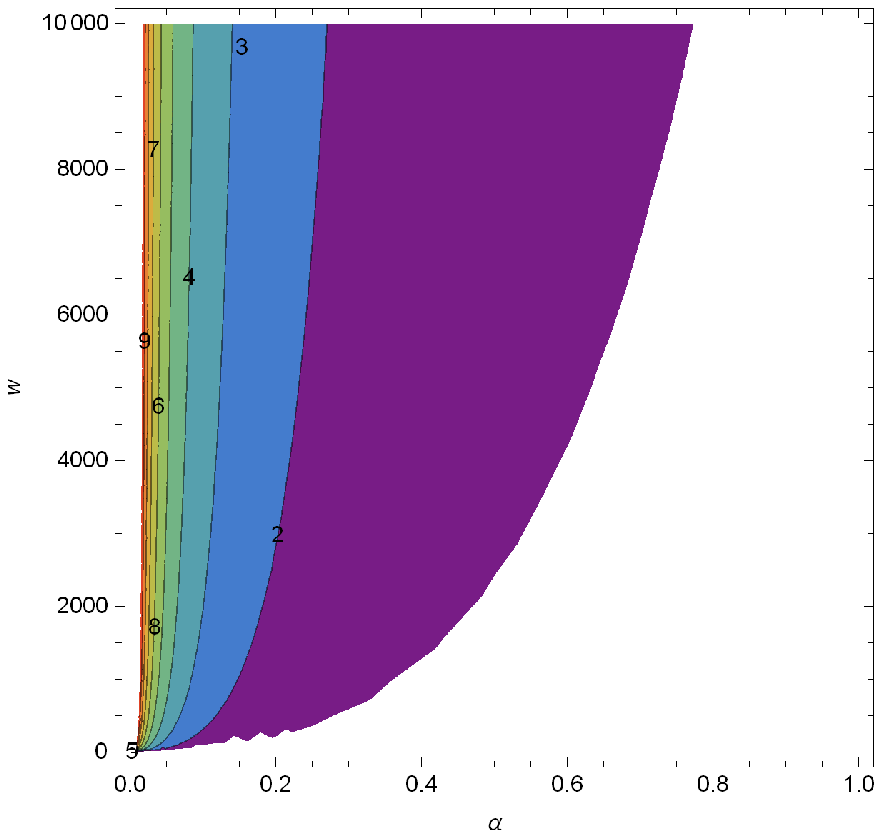}
\includegraphics[width=18pc]{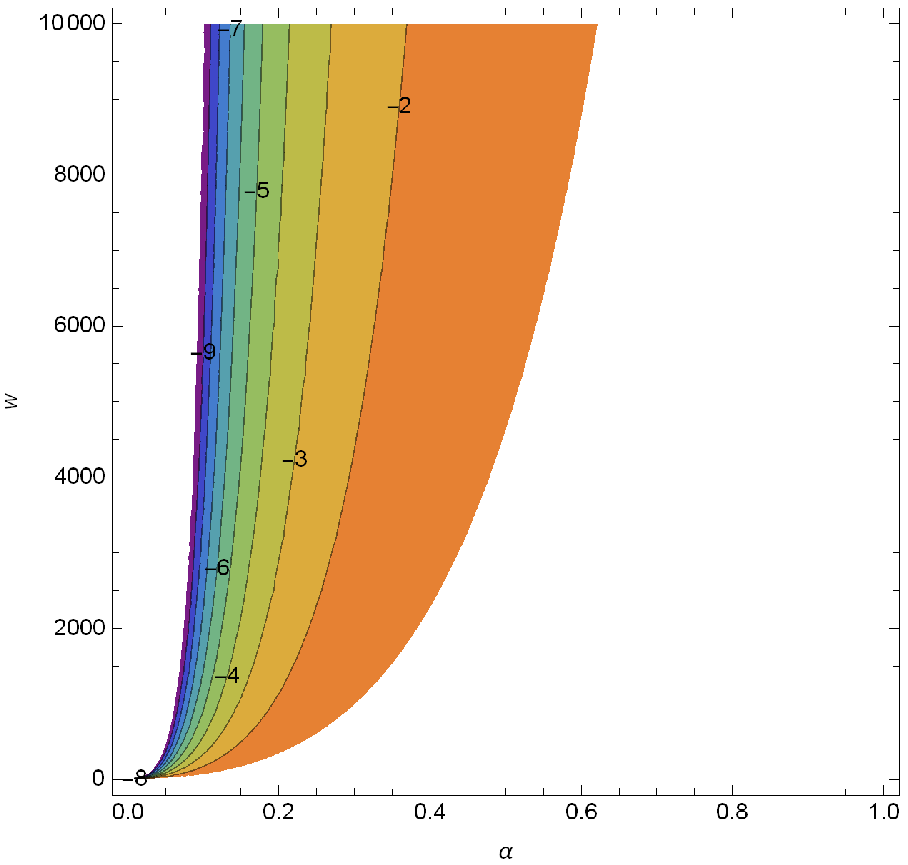}
\caption{Second and third Swampland criteria analysis for the
E-model (N=2). Contour plots of $V'(\phi)/V(\phi)$(left plot) and
of $-V''(\phi)/V(\phi)$(right plot), evaluated at the first
horizon crossing for the ranges of values of the free parameters
$\alpha = [0 , 1]$, $w = [0.01, 10000]$  in reduced Planck units
for $N=60$.} \label{e2sw23}
\end{figure}

Next we shall consider the E-model potential (N=2). For this
potential, the first Swampland criterion assumes the form (from
now on we set $\kappa = 1$ without further notice),
\begin{equation}
    \centering\label{emod df}
    \Delta \phi = \sqrt{\frac{3 w}{2}}\ln\frac{1 + \frac{4 \sqrt{\alpha}}{ \sqrt{3 w}}}{1 + \frac{4 \sqrt{\alpha}}{ \sqrt{3 w}}+ \frac{8 N \alpha}{3 w}},
\end{equation}
the contour plot of which for $N = 60$ is given in Fig. \ref{e2sw1}.
From Fig. \ref{e2sw1}, we see that (\ref{S.C. deltaphi}) is
verified. Although, in order not to jump into conclusions, we have
to check the other two conjectures as well. So, the second
Swampland criterion while also taking (\ref{S.C. v'/v}) into
consideration, becomes,
\begin{equation}
    \centering\label{emod lamb}
    \frac{V'(\phi_i)}{V(\phi_i)} = 4 \sqrt{\frac{2}{3}}\frac{1}{1 + \frac{4 \sqrt{\alpha}}{ \sqrt{3 w}}+ \frac{8 N \alpha}{3 w}} \frac{\frac{1}{\sqrt{w}}}{1 - \frac{1}{1 + \frac{4 \sqrt{\alpha}}{\sqrt{3 w}}+ \frac{8 N \alpha}{3 w}}},
\end{equation}
the contour plot of which for $N = 60$ we depict in Fig. \ref{e2sw23}.
From Fig. \ref{e2sw23}, we see that (\ref{S.C. v'/v}) is fairly
satisfied as well. Last thing to check is the third Swampland
criterion, which in this particular case assumes the following
form:
\begin{equation}
    \centering\label{emod c2}
    -\frac{V''(\phi_i)}{V(\phi_i)} =\frac{8}{3 w} \frac{\frac{4 \sqrt{\alpha}}{\sqrt{3 w}} + \frac{8 N \alpha}{3 w} -3}{3 w\left(\frac{4 \sqrt{\alpha}}{\sqrt{3 w}} + \frac{8 N \alpha}{3 w}\right)^2}.
\end{equation}
Substituting $N = 60$ in (\ref{emod c2}) and taking (\ref{S.C.
v''/v}) into account, we obtain its respective contour plot in
Fig. \ref{e2sw23}.

Looking at Fig. \ref{e2sw23} and taking into account Fig.
\ref{e2sw1} as well, we deduce that all Swampland criteria are
satisfied for the following values of the parameters:
\begin{equation}
\centering\label{emod const}
\boxed{
\begin{array}{rcl}
\alpha&=&[0 , 0.52]\\
w&=&[0.01 , 10000]
\end{array}}
\end{equation}

\subsection{Swampland Criteria for the Hilltop quadratic model}

\begin{figure}
\centering
\includegraphics[width=18pc]{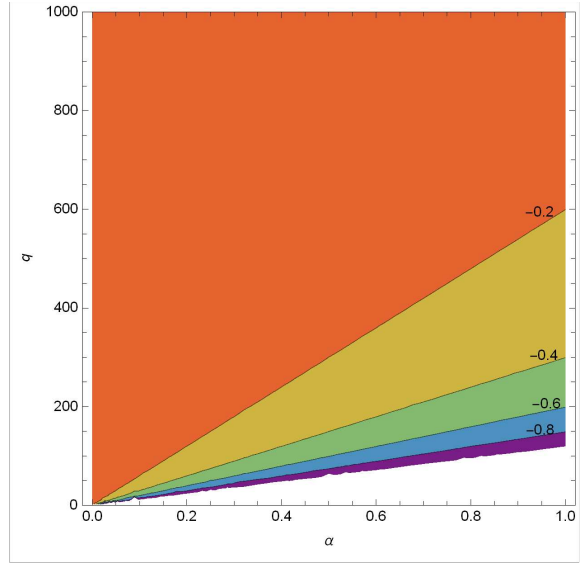}
\caption{First Swampland criterion analysis for the Hilltop
quadratic model. Contour plot of $\Delta\phi$, evaluated at the
first horizon crossing for the ranges of values of the free
parameters $\alpha = [0, 1]$, $q = [2 , 1000]$ in reduced Planck
units for $N=60$.} \label{hillsw1}
\end{figure}

As the next potential, we consider the Hilltop quadratic model.
Now, the first Swampland criterion, taking into consideration
(\ref{S.C. deltaphi}), takes the form,
\begin{equation}
    \centering\label{hill df}
    \Delta \phi = \sqrt{\alpha + q^2 + \sqrt{\alpha(\alpha + 2 q^2)}} - \sqrt{\alpha + 4 N \alpha^2 + q^2 + \sqrt{\alpha(\alpha + 2 q^2)}},
\end{equation}
the contour plot of which for $N = 60$ is given in Fig. \ref{hillsw1}.

\begin{figure}
\centering
\includegraphics[width=18pc]{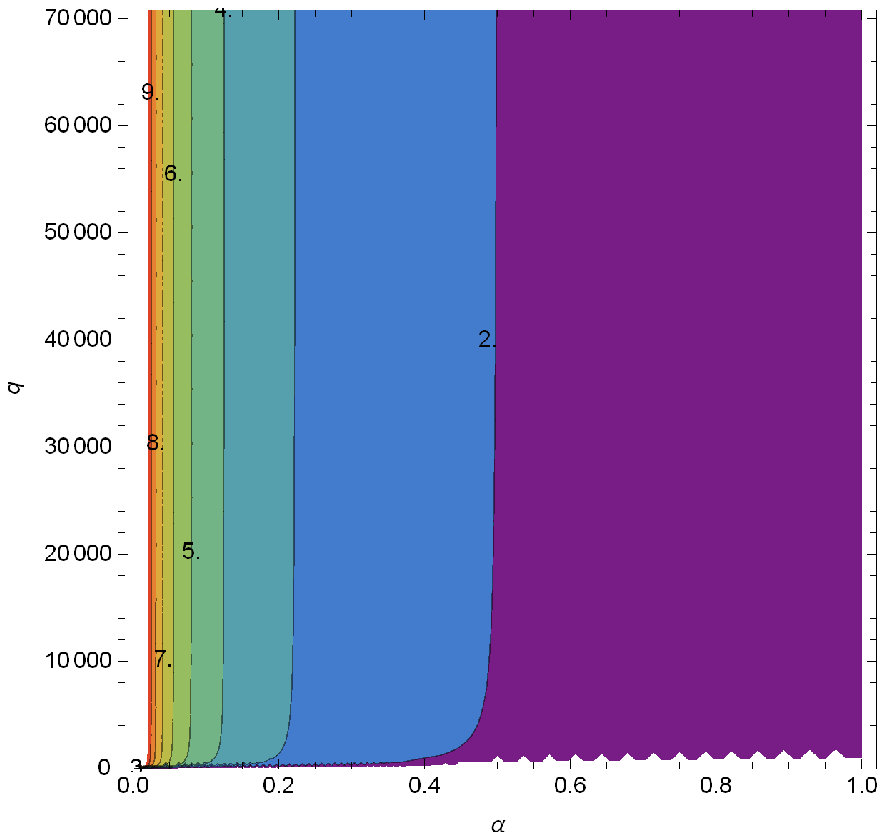}
\includegraphics[width=18pc]{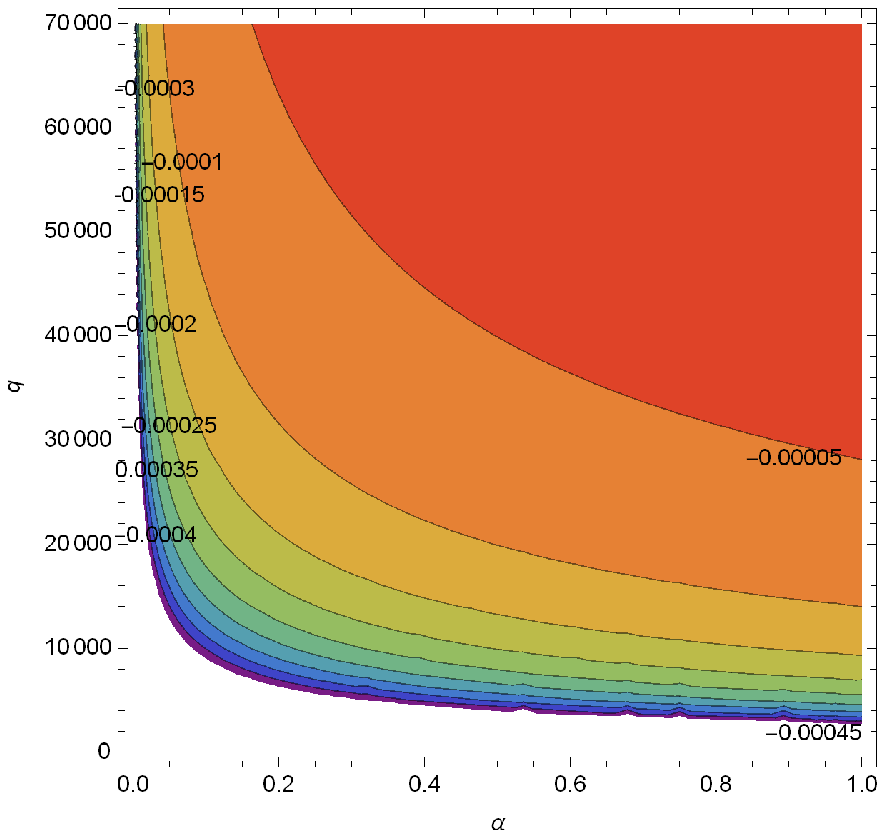}
\caption{Second and third Swampland criteria analysis for the Hilltop Quadratic model. Contour plots of $V'(\phi)/V(\phi)$(left plot) and of $-V''(\phi)/V(\phi)$(right plot), evaluated at the first horizon crossing for the ranges of values of the free parameters $\alpha = [0, 1]$, $q = [2 , 70000]$(left plot) and $\alpha = [0, 1],$ $q = [2 , 70000]$(right plot) in reduced Planck units for $N=60$.}
\label{hillsw23}
\end{figure}

From Fig. \ref{hillsw1}, we see that (\ref{S.C. deltaphi}) is
valid only in a specific region. We proceed to the verification of
the other two criteria. Concretely, the second Swampland
criterion, bearing in mind (\ref{S.C. v'/v}), becomes
\begin{equation}
    \centering\label{hill lamb}
    \frac{V'(\phi_i)}{V(\phi_i)} = 2\frac{\sqrt{\alpha + 4 N \alpha + q^2 + \sqrt{\alpha(\alpha + 2 q^2)}}}{\alpha + 4 N \alpha +q^2 + \sqrt{\alpha(\alpha + 2 q^2)}},
\end{equation}
the contour plot of which for $N = 60$ is illustrated in
Fig. \ref{hillsw23}.
From Fig. \ref{hillsw23}, we see that (\ref{S.C. v'/v}) is fairly
satisfied for all $q$ but restricts severely $\alpha$. Lastly, we
check the third Swampland criterion, which in this particular case
assumes the following form:
\begin{equation}
    \centering\label{hill c2}
    -\frac{V''(\phi_i)}{V(\phi_i)} = -\frac{2}{\alpha + 4 N \alpha +q^2 + \sqrt{\alpha(\alpha + 2 q^2)}}.
\end{equation}
Substituting $N = 60$ in (\ref{hill c2}) and taking into account
(\ref{S.C. v''/v}), we obtain its contour plot which is shown in
Fig. \ref{hillsw23}. By looking at Fig.\ref{hillsw23}, it is
evident that (\ref{S.C. v''/v}) is not satisfied regardless of the
values $\alpha$ or $q$ may assume. Therefore, we reach the
conclusion that the Swampland criteria for the Hilltop quadratic
potential are not satisfied.

\section{Conclusions}

In this work we studied a class of $f(R,\phi)$ gravity models
which can effectively lead to a rescaled Einstein-Hilbert
canonical scalar field gravity in the large curvature regime.
These effective $f(R,\phi)$ theories result to gravitational
models for which the standard curvature term of the
Einstein-Hilbert Lagrangian is replaced by $\alpha R$, where the
dimensionless parameter $\alpha$ takes values in the range
$0<\alpha<1$. The main motivation for studying these effective
$f(R,\phi)$ theories is the compatibility of the resulting
rescaled Einstein-Hilbert scalar field theory with the Swampland
criteria. In this new theoretical framework we studied several
models appearing in the latest Planck collaboration release
\cite{Akrami:2020zfz}, focusing on the simultaneous compatibility
of the models with the Planck inflationary data and the Swampland
criteria. As we demonstrated, some models cannot be compatible
with the Planck constraints on inflation, even for the rescaled
Einstein-Hilbert framework, however in some cases it is possible
to have models compatible with the Planck constraints and with the
Swampland criteria. As we showed it is possible to obtain
compatibility between the constraints that the latest 2018 Planck
data imply for the primordial tilt and the tensor-to-scalar ratio
and the limitations that the Swampland criteria introduce. For the
models that this simultaneous compatibility with the Planck
constraints and the Swampland criteria occurs, the parameter
$\alpha$ must take values much smaller from unity.

\end{document}